\documentclass[twocolumn]{aastex63}

\usepackage{mathrsfs}

\shorttitle{SHELLQs. XXIV. 54 New Quasars and Candidate Obscured Quasars}
\shortauthors{Matsuoka et al.}

\begin{document}

\title{Subaru High-$z$ Exploration of Low-Luminosity Quasars (SHELLQs). XXIV. 
54 New Quasars and Candidate Obscured Quasars at $5.71 \le z \le 7.02$}

\correspondingauthor{Yoshiki Matsuoka}
\email{matsuoka.yoshiki.ld@ehime-u.ac.jp}


\author[0000-0001-5063-0340]{Yoshiki Matsuoka}
\affil{Research Center for Space and Cosmic Evolution, Ehime University, Matsuyama, Ehime 790-8577, Japan.}

\author[0000-0002-4923-3281]{Kazushi Iwasawa}
\affil{ICREA and Institut de Ci{\`e}ncies del Cosmos, Universitat de Barcelona, IEEC-UB, Mart{\'i} i Franqu{\`e}s, 1, 08028 Barcelona, Spain.}

\author[0000-0003-2984-6803]{Masafusa Onoue}
\affiliation{Waseda Institute for Advanced Study (WIAS), Waseda University, Shinjuku, Tokyo 169-0051, Japan.}
\affiliation{Kavli Institute for the Physics and Mathematics of the Universe, WPI, The University of Tokyo, Kashiwa, Chiba 277-8583, Japan.}

\author[0000-0001-9452-0813]{Takuma Izumi}
\affil{National Astronomical Observatory of Japan, Mitaka, Tokyo 181-8588, Japan.}

\author[0000-0002-0106-7755]{Michael A. Strauss}
\affil{Department of Astrophysical Sciences, Princeton University, Peyton Hall, Princeton, NJ 08544, USA.}

\author[0000-0002-2651-1701]{Masayuki Akiyama}
\affil{Astronomical Institute, Tohoku University, Aoba, Sendai, 980-8578, Japan.}

\author[0000-0003-4569-1098]{Kentaro Aoki}
\affil{Subaru Telescope, National Astronomical Observatory of Japan, Hilo, HI 96720, USA.}

\author[0009-0007-0864-7094]{Junya Arita}
\affiliation{Department of Astronomy, School of Science, The University of Tokyo, Tokyo 113-0033, Japan.}

\author[0000-0001-8917-2148]{Xuheng Ding}
\affil{School of Physics and Technology, Wuhan University, Wuhan 430072, China.}
\affil{Kavli Institute for the Physics and Mathematics of the Universe, WPI, The University of Tokyo, Kashiwa, Chiba 277-8583, Japan.}

\author[0000-0001-6186-8792]{Masatoshi Imanishi}
\affil{National Astronomical Observatory of Japan, Mitaka, Tokyo 181-8588, Japan.}
\affil{Department of Astronomical Science, Graduate University for Advanced Studies (SOKENDAI), Mitaka, Tokyo 181-8588, Japan.}

\author[0000-0003-3954-4219]{Nobunari Kashikawa}
\affil{Department of Astronomy, School of Science, The University of Tokyo, Tokyo 113-0033, Japan.}

\author[0000-0002-3866-9645]{Toshihiro Kawaguchi}
\affil{Graduate School of Science and Engineering, University of Toyama, Toyama, Toyama 930-8555, Japan.}

\author[0000-0003-3214-9128]{Satoshi Kikuta}
\affil{Department of Astronomy, School of Science, The University of Tokyo, Tokyo 113-0033, Japan.}

\author[0000-0002-4052-2394]{Kotaro Kohno}
\affil{Institute of Astronomy, The University of Tokyo, Mitaka, Tokyo 181-0015, Japan.}
\affil{Research Center for the Early Universe, University of Tokyo, Tokyo 113-0033, Japan.}

\author[0000-0003-1700-5740]{Chien-Hsiu Lee}
\affil{Hobby-Eberly Telescope, McDonald Observatory, Fort Davis, TX 79734, USA.}

\author[0000-0002-7402-5441]{Tohru Nagao}
\affil{Research Center for Space and Cosmic Evolution, Ehime University, Matsuyama, Ehime 790-8577, Japan.}

\author[0000-0002-2099-0254]{Camryn L. Phillips}
\affil{Department of Astrophysical Sciences, Princeton University, Peyton Hall, Princeton, NJ 08544, USA.}

\author[0009-0003-5438-8303]{Mahoshi Sawamura}
\affil{Department of Astronomy, School of Science, The University of Tokyo, Tokyo 113-0033, Japan.}
\affil{National Astronomical Observatory of Japan, Mitaka, Tokyo 181-8588, Japan.}

\author[0000-0002-0000-6977]{John D. Silverman}
\affiliation{Kavli Institute for the Physics and Mathematics of the Universe, WPI, The University of Tokyo, Kashiwa, Chiba 277-8583, Japan.}
\affiliation{Department of Astronomy, School of Science, The University of Tokyo, Tokyo 113-0033, Japan.}
\affiliation{Center for Data-Driven Discovery, Kavli IPMU (WPI), UTIAS, The University of Tokyo, Kashiwa, Chiba 277-8583, Japan}
\affiliation{Center for Astrophysical Sciences, Department of Physics \& Astronomy, Johns Hopkins University, Baltimore, MD 21218, USA}\

\author[0000-0003-3769-6630]{Ayumi Takahashi}
\affil{National Astronomical Observatory of Japan, Mitaka, Tokyo 181-8588, Japan.}

\author[0000-0002-3531-7863]{Yoshiki Toba}
\affiliation{Department of Physical Sciences, Ritsumeikan University, Kusatsu, Shiga 525-8577, Japan.}
\affiliation{Academia Sinica Institute of Astronomy and Astrophysics, Taipei 10617, Taiwan.}
\affiliation{Research Center for Space and Cosmic Evolution, Ehime University, Matsuyama, Ehime 790-8577, Japan.}



\begin{abstract}

We present spectroscopic identification of 43 quasars and 11 candidate obscured quasars in the epoch of reionization (EoR) at $5.71 \le z \le 7.02$, along with 
29 galaxies at similar redshifts.
This is the 24th publication from the Subaru High-$z$ Exploration of Low-Luminosity Quasars (SHELLQs) project, which exploits the Hyper Suprime-Cam (HSC)
Subaru Strategic Program (SSP) imaging survey to search for EoR quasars.
The HSC-SSP survey has completed, and this paper is likely the final installment of major (unobscured) quasar discoveries from the SHELLQs project.
In addition to the EoR objects, we identified five strong [\ion{O}{3}] line emitters at $z < 1$, 30 Galactic brown dwarfs, and 14 passive galaxies at $z \sim 2$, which contaminated our sample of photometric quasar candidates.
The present paper focuses on describing the immediate outcome of the spectroscopic observations, while a statistical analysis of the full SHELLQs sample will be presented in our
next publication.

\bigskip
\bigskip
\bigskip

\end{abstract}




\section{Introduction} \label{sec:intro}

For more than 10 years, we have been carrying out a project to search for and characterize low-luminosity quasars in the epoch of reionization (EoR, referring to $z > 5.6$ in this paper).
The project, ``Subaru High-$z$ Exploration of Low-Luminosity Quasars \citep[SHELLQs;][]{p1}", is based on the Subaru Strategic Program \citep[SSP;][]{aihara18}
survey with Hyper Suprime-Cam \citep{miyazaki18}, a cutting-edge wide-field imager mounted at the prime focus of the Subaru Telescope.
Candidate EoR quasars were selected from the HSC images, mainly utilizing the $i$-, $z$-, and $y$-band photometry, with a Bayesian probabilistic algorithm \citep{mortlock12} or simple color cuts.
Spectroscopic identification of the candidates has been performed with 
the Subaru Telescope and the Gran Telescopio Canarias (GTC).

We have so far reported discovery of 139 broad-line (unobscured) quasars at $5.6 < z < 7.1$ 
in a series of publications \citep{p1,p2,p4,p10,p16}.
The identified objects include the first known low-luminosity quasar at $z > 7$ \citep{p7} and the first confirmed pair of quasars at $z > 6$ \citep{p20}.
The established sample was used to measure the quasar luminosity functions at $z = 6$ \citep{p5} and at $z = 7$ \citep{p19}, which indicated
that quasars contribute only in a minor way 
to cosmic reionization.
The black hole (BH) masses and Eddington ratios were estimated from the follow-up near-infrared (IR) spectroscopy \citep{onoue19, onoue21} or from the spectral shape
around Ly$\alpha$ compared to low-$z$ quasars \citep{takahashi24}.
Observations with the Atacama Large Millimeter/submillimeter Array (ALMA) detected [\ion{C}{2}] 158-$\mu$m line and far-IR continuum in many of the quasars \citep{izumi18,izumi19,izumi21a,izumi21b,sawamura25}, and revealed that the ratios between the host dynamical mass and BH mass are more or less consistent
with the stellar-to-BH mass ratios measured in galaxies at $z \sim 0$.

Before the James Webb Space Telescope \citep[JWST;][]{rigby23} was launched, the SHELLQs quasars served as a unique statistical sample of the lowest-luminosity
quasars or active galactic nuclei (AGNs) known in the EoR.
Now with data from JWST, the SHELLQs sample is taking on a new role.
\citet{ding23, ding25} observed 12 SHELLQs quasars with NIRCam, and achieved the first secure detection of starlight in hosts of EoR quasars.
The same quasars were also observed with NIRSpec, allowing us to estimate BH masses as well as the star formation history in some of the objects with stellar absorption lines \citep[][M. Onoue et al. 2025, in preparation]{onoue24}.
Related studies of the stellar-to-BH mass relation and the extended spectral properties are ongoing (e.g., J. D. Silverman et al. 2025, in preparation; C. L. Phillips et al. 2025, in preparation).
In the meantime, JWST has revealed an abundant population of candidate low-luminosity AGNs in the EoR, which manifest as galaxies with broad Balmer lines or as ``little red dots" \citep[e.g.,][]{onoue23, harikane23, kocevski23, kocevski24, matthee24, greene24}.
The SHELLQs quasars occupy a unique luminosity range bridging those new AGN candidates and classical quasars, and may represent the missing link between the two populations \citep{matsuoka25}.


The HSC-SSP survey completed its 330 nights of observation in 2021 December. 
The SHELLQs project continued spectroscopic identification of quasar candidates through 2025 March, and has now covered all primary candidates from the entire HSC-SSP dataset.
This paper presents the results from all the spectroscopic observations since our last discovery paper \citep{p16}.
This is likely the final installment of major (unobscured) quasar discoveries from the project, although discoveries of other populations, such as obscured quasars, may still
continue \citep[e.g.,][]{kato20, iwamoto25, matsuoka25}.
We plan to analyze the statistics of the full SHELLQs sample in the next paper (Y. Matsuoka et al. 2025, in preparation), and thus the present paper 
is kept short and concise.
The sample and spectroscopic observations are presented in \S \ref{sec:obs}, and the results and short discussions appear in \S \ref{sec:results}.
A summary is given in \S \ref{sec:summary}.
All magnitudes refer to point-spread function (PSF) magnitudes presented in the AB system \citep{oke83} unless otherwise noted, corrected for
Galactic extinction \citep{schlegel98}. 
We refer to $z$-band magnitudes with the AB subscript (``$z_{\rm AB}$"), while redshift $z$ appears without a subscript. 
The cosmological parameters are assumed to be $H_0 = 70$ km s$^{-1}$ Mpc$^{-1}$, $\Omega_{\rm M} = 0.3$, and $\Omega_{\rm \Lambda} = 0.7$.

\section{Observations} \label{sec:obs}

The candidates presented here were selected from the S21A and S23B internal data releases (DRs) of the HSC-SSP.
The latter is equivalent to the final public DR of the survey, which will take place in the near future.
In \citet{p16} and earlier papers, quasar candidates were limited to those with $\mu$/$\mu_{\rm PSF}$ $<$ 1.2 in the $z$ band (for $i$-band dropouts, which are
$z \sim 6$ quasar candidates) or with $\mu$/$\mu_{\rm PSF}$ $<$ 3.0 in the $y$ band (for $z$-band dropouts, $z \sim 7$ quasar candidates).
Here, $\mu$ and $\mu_{\rm PSF}$ represent the image adaptive moments of a source and its PSF model, respectively.
We have tested a relaxed shape condition since then, and as a result, the present paper includes more extended $i$-dropouts with $\mu$/$\mu_{\rm PSF}$ $<$ 3.0 in the $z$ band.
All the other selection criteria are identical to those described in our past publications \citep[e.g.,][]{p16}.
We consistently applied updated selection criteria to all previous imaging data (reprocessed with the latest pipeline), and as a result, the current candidates are distributed across
the entire 1100-deg$^2$ area of the completed HSC-SSP footprint.

Table \ref{tab:obsjournal} in the Appendix presents the journal of spectroscopic observations for a total of 132 candidates.
At the Subaru Telescope, we used the Faint Object Camera and Spectrograph \citep[FOCAS;][]{kashikawa02} to cover the red-optical wavelengths
from 0.75 to 1.0 $\mu$m.
We used the VPH900 grism and 1\arcsec.0 slit, which provide a spectral resolution of $R \sim 1200$.
All the observations have been carried out in the classical mode, with the program IDs of S18B-011I, S22A-025, S22B-042, S24A-053, and S24B-025.
At the GTC, all the observations have been conducted in the queue mode.
The program IDs are GTC42-21B, GTC30-22A, GTC23-24A, and GTC3-24B.
We used the Optical System for Imaging and low-intermediate-Resolution Integrated Spectroscopy \citep[OSIRIS;][]{cepa00}.
The R2500I grism was used with the 1\arcsec.0 longslit, giving a spectral coverage from 0.75 to 1.0 $\mu$m at a resolution of $R \sim 1500$.

\section{Results and Discussion} \label{sec:results}

\begin{figure*}
\epsscale{0.97}
\plotone{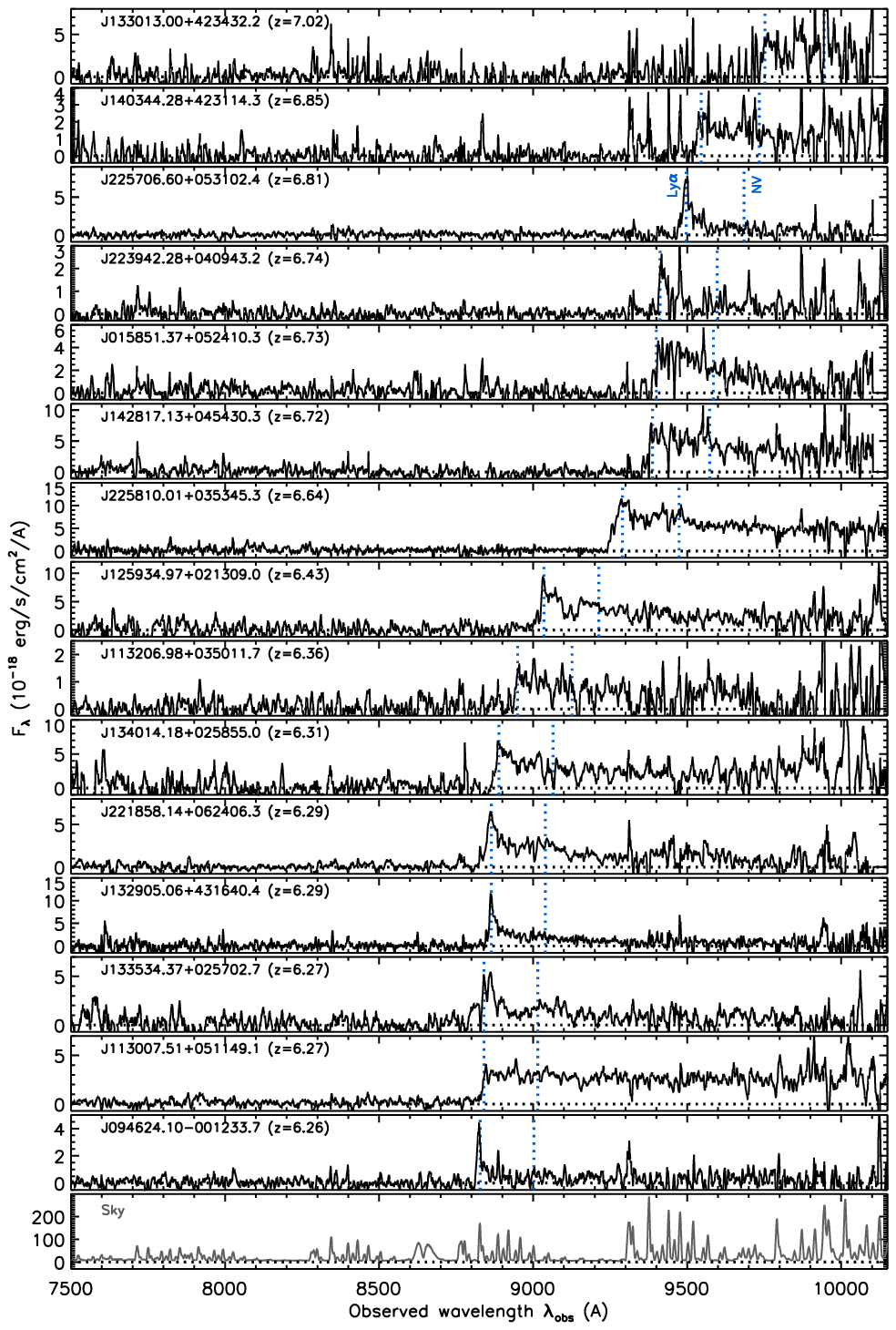}
\caption{Discovery spectra of the first set of 15 quasars, displayed in decreasing order of redshift.
The object name and the estimated redshift \citep[with uncertainty $\Delta z \sim 0.01 - 0.1$;][]{p16} are indicated at the top left corner of each panel.
The blue dotted lines mark the expected positions of the Ly$\alpha$ and \ion{N}{5} $\lambda$1240 emission lines, given the redshifts.
The spectra were smoothed using inverse-variance weighted means over 3 -- 9 pixels (depending on the signal-to-noise ratio), for display purposes.
The bottom panel displays a sky spectrum, as a guide to the expected noise.
\label{fig:spec1}}
\end{figure*}

\begin{figure*}
\epsscale{1.0}
\plotone{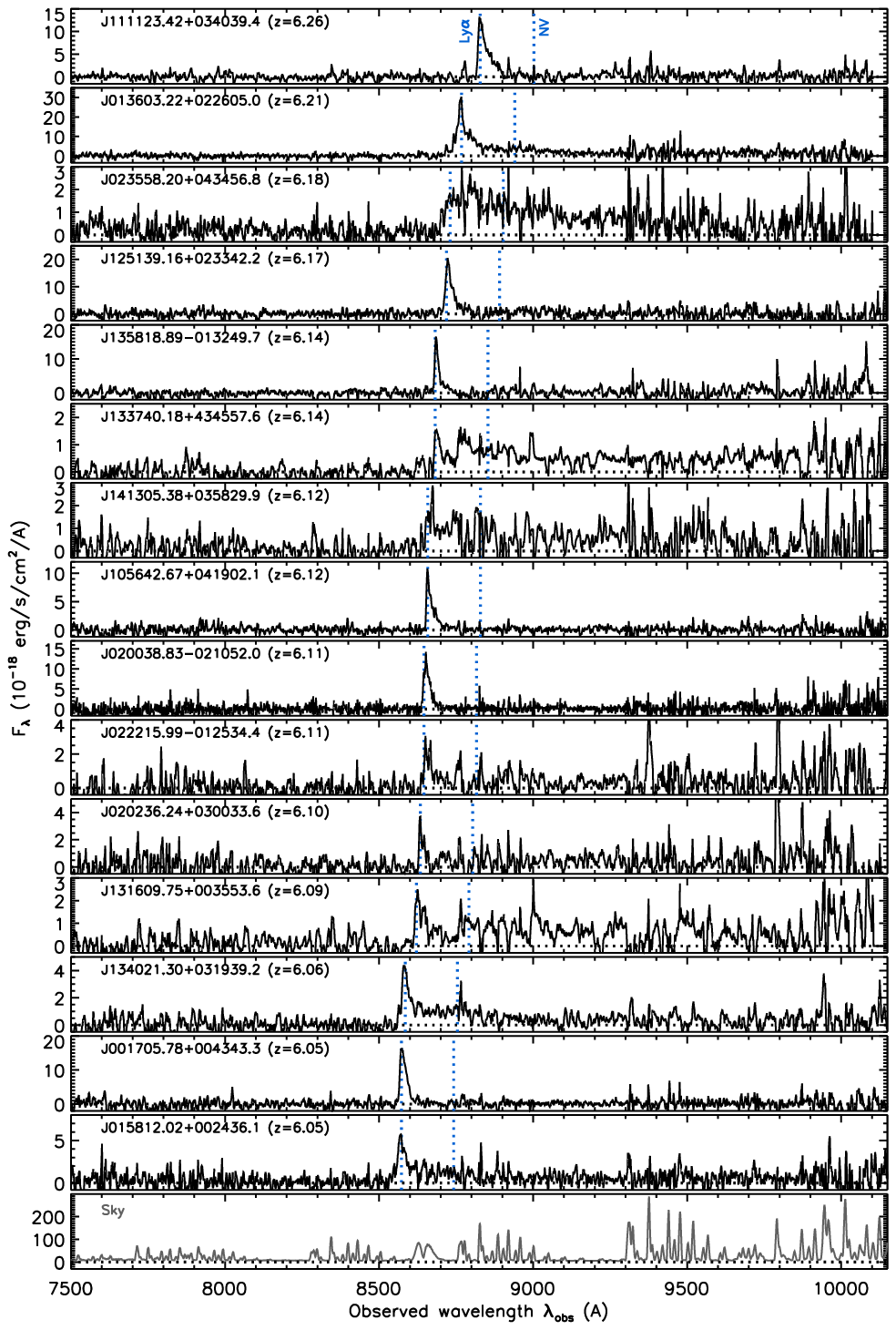}
\caption{Same as Figure \ref{fig:spec1}, but for the second set of 15 quasars.
\label{fig:spec2}}
\end{figure*}

\begin{figure*}
\epsscale{1.0}
\plotone{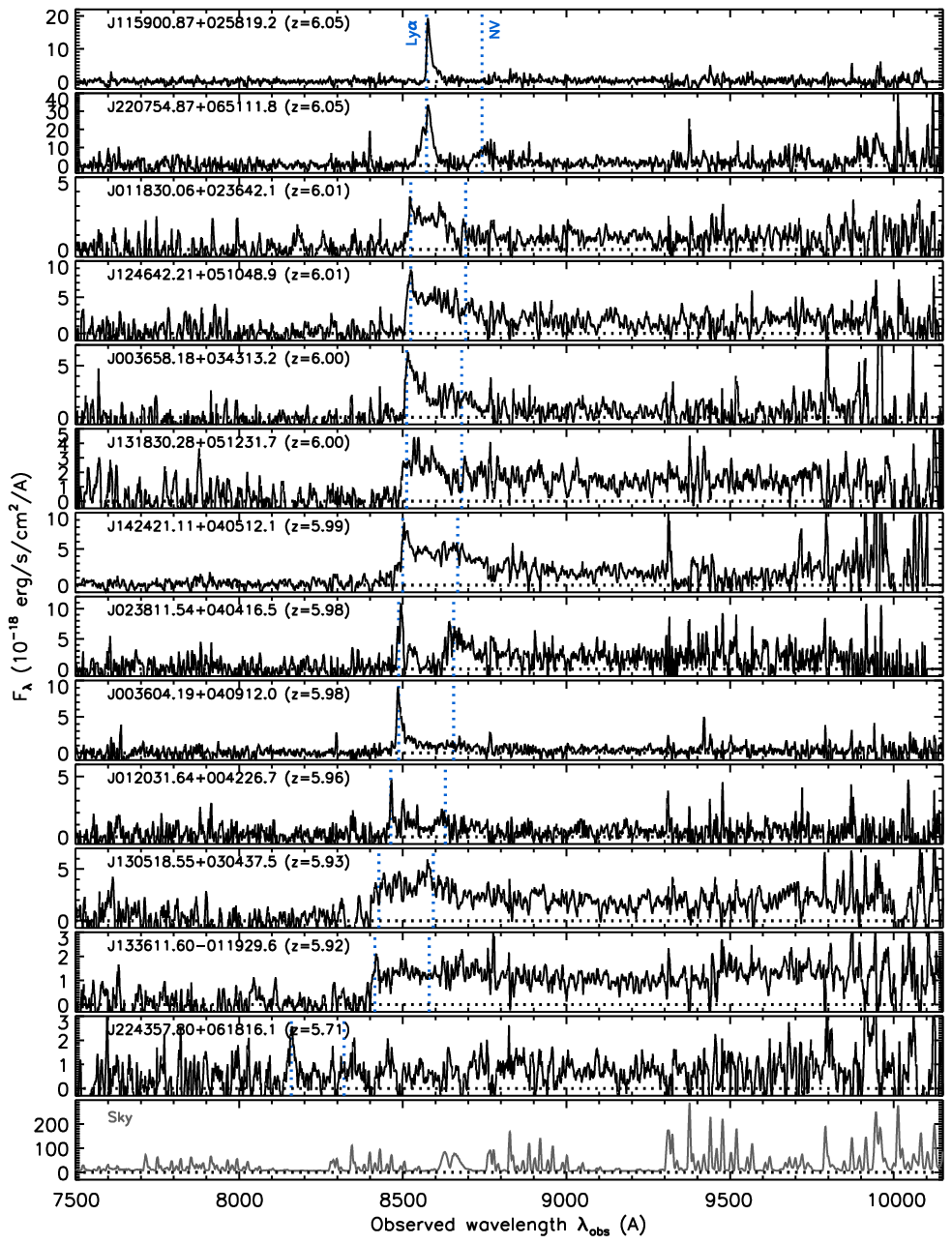}
\caption{Same as Figure \ref{fig:spec1}, but for the third set of 13 quasars.
\label{fig:spec3}}
\end{figure*}

\begin{figure*}
\epsscale{1.0}
\plotone{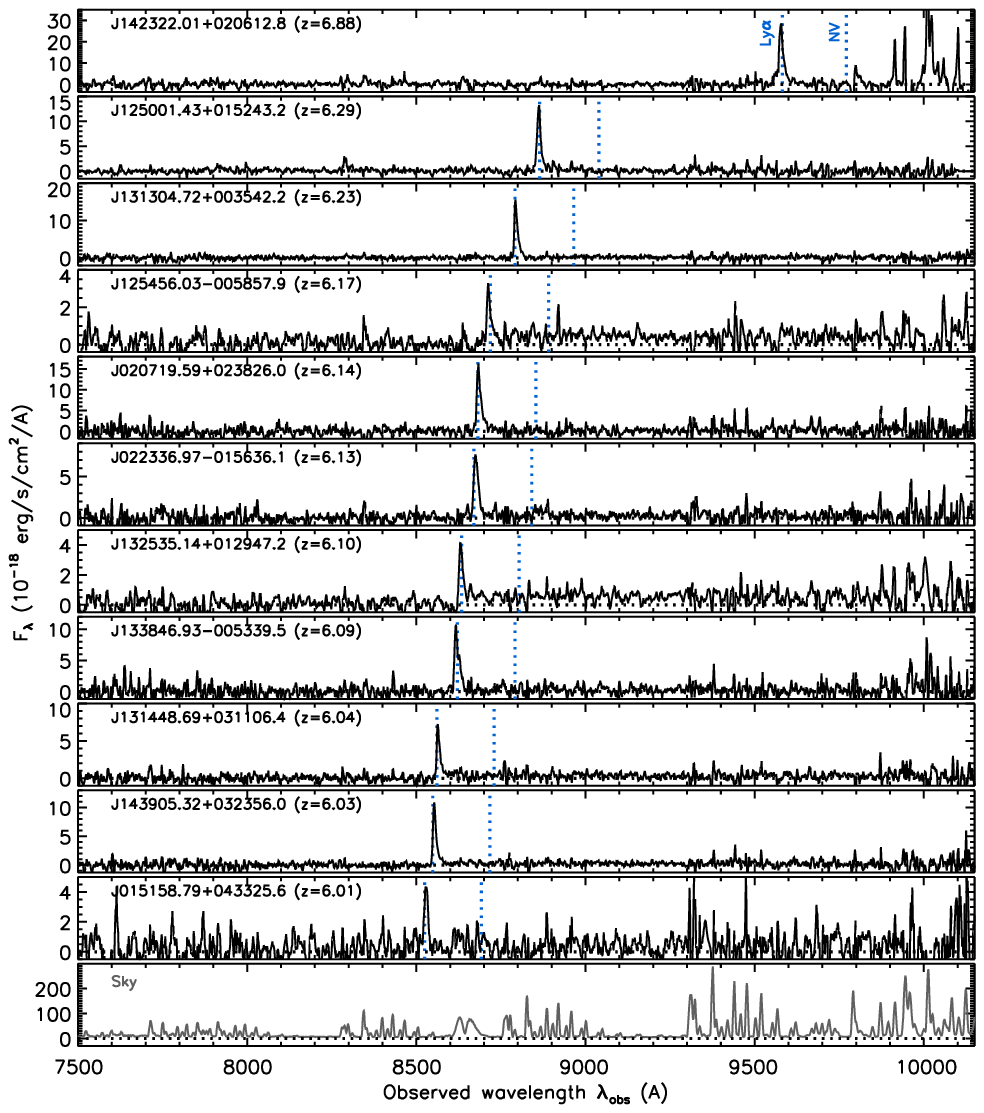}
\caption{Same as Figure \ref{fig:spec1}, but for 11 candidate obscured (narrow-line) quasars.
\label{fig:spec4}}
\end{figure*}

Figures \ref{fig:spec1} -- \ref{fig:spec3} present the reduced spectra of 43 candidates that turned out to be quasars at $z = 5.71 - 7.02$.
They are characterized by broad Ly$\alpha$ line and/or blue continuum emission, as well as the sharp spectral break due to the \ion{H}{1} absorption by the intergalactic medium.
$J142817.13+045430.3$ was independently discovered and reported by \citet{belladitta25}.
Some of the spectra are severely affected by the noise from sky emission, especially at the redder part, where there are strong sky lines and the detector sensitivity is relatively low.
The spiky features at 9300 -- 9500 \AA\ in $J133013.00+423432.2$ and $J140344.28+423114.3$ are due to sky noise.
We also found 11 high-$z$ objects with very luminous and narrow Ly$\alpha$ emission (Figure \ref{fig:spec4}), with line luminosities $L_{\rm Ly\alpha} > 10^{43}$ erg s$^{-1}$ and full widths at half maximum (FWHMs) $<$ 500 km s$^{-1}$ after correcting for the instrumental resolution.
Our past publications \citep{p1, p2, p4, p10, p16} have consistently classified such objects as ``candidate obscured (narrow-line) quasars", since Ly$\alpha$ emitters at 
lower redshifts are known to be dominated by AGNs at $L_{\rm Ly\alpha} > 10^{43}$ erg s$^{-1}$ \citep[e.g.,][]{konno16, sobral18, spinoso20}.
Indeed, we observed 11 targets (out of 34 known at that time) from this population with JWST/NIRSpec in Cycles 1 and 2, and found that seven of them exhibit broad \ion{H}{1} Balmer lines and \ion{He}{1} lines, indicating that they are mildly-obscured quasars \citep{matsuoka25}.
$J020719.59+023826.0$ and $J142322.01+020612.8$ reported in this paper were included among the JWST targets, and both of them have broad rest-optical lines.

\begin{figure*}
\epsscale{1.0}
\plotone{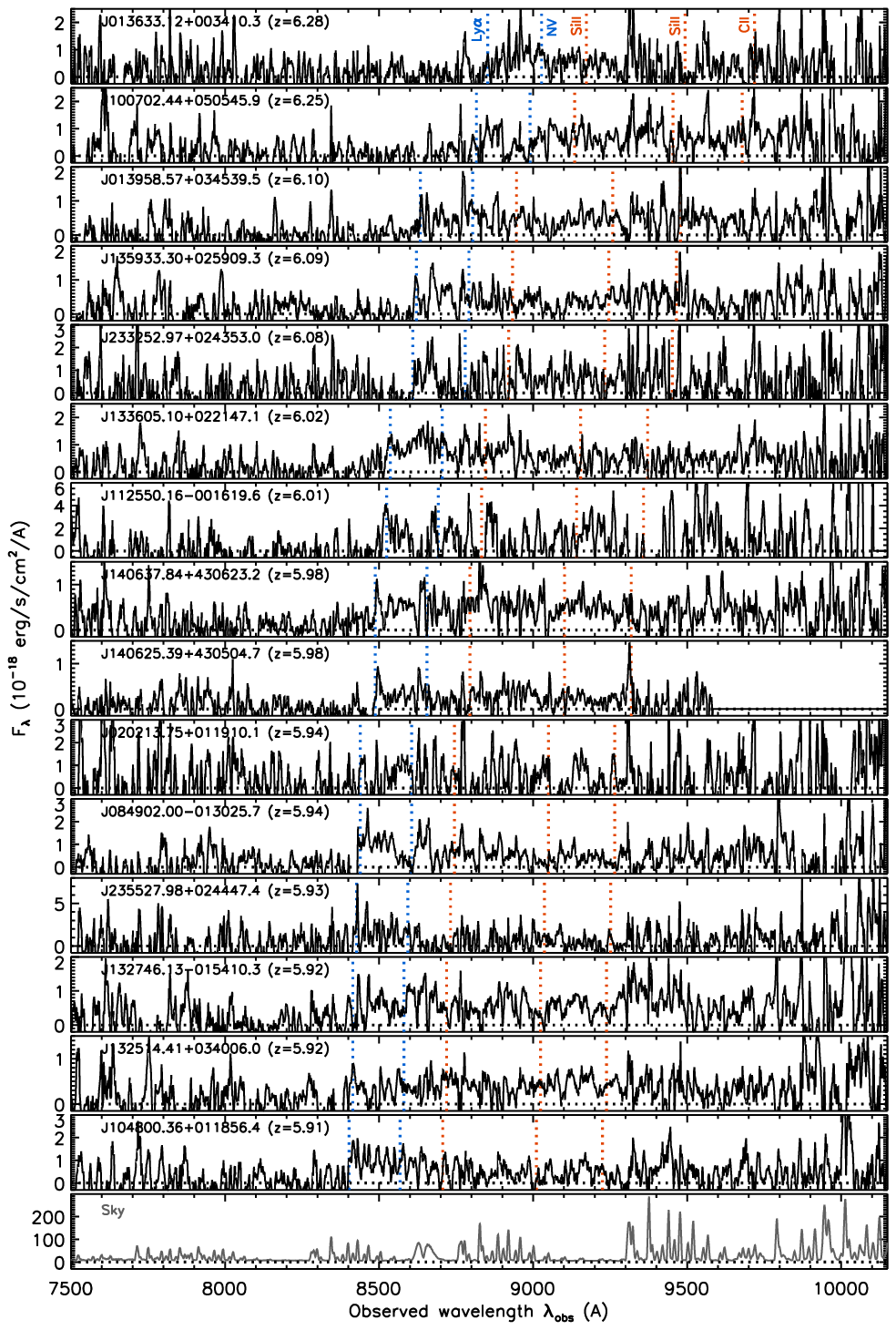}
\caption{Same as Figure \ref{fig:spec1}, but for the first set of 15 galaxies.
The expected positions of \ion{Si}{2} $\lambda$1260, \ion{Si}{2} $\lambda$1304, and \ion{C}{2} $\lambda$1335 are marked by the red dotted lines.
The flux excess at $\sim$8780 \AA\ in $J013633.12+003410.3$ is due to sky noise (see the bottom panel).
\label{fig:spec5}}
\end{figure*}

\begin{figure*}
\epsscale{1.0}
\plotone{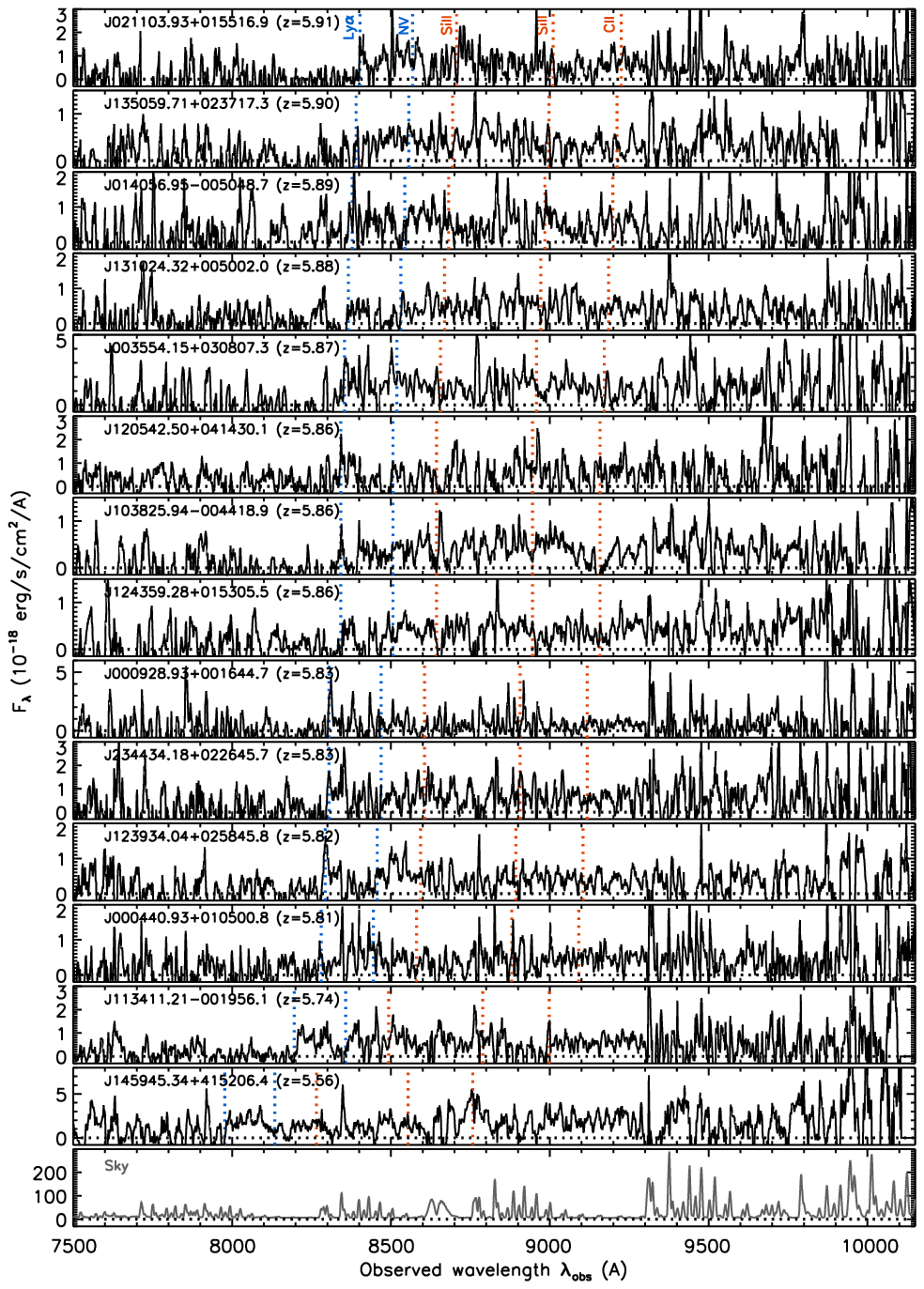}
\caption{Same as Figure \ref{fig:spec5}, but for the second set of 14 galaxies.
\label{fig:spec6}}
\end{figure*}

Galaxies at similar redshifts without strong Ly$\alpha$ emission ($L_{\rm Ly\alpha} < 10^{43}$ erg s$^{-1}$) are presented in Figures \ref{fig:spec5} -- \ref{fig:spec6}.
Among them, $J140637.84+430623.2$ and $J140625.39+430504.7$ were found within a single circular FOCAS field-of-view with a diameter of 6 arcmin.
The galaxies are at the identical redshift of $z = 5.98$, and
their angular separation (2.6 arcmin) corresponds to 900 kpc in projected proper distance.
The presence of two such luminous galaxies ($M_{1450} = -23.3$ and $-22.3$) in close proximity suggests the possibility of a galaxy overdensity in this region.
Additional spectroscopy of surrounding fainter galaxies would be required to test this possibility \citep[e.g.,][]{toshikawa12}.

We note that the separation between quasars and galaxies is not always straightforward, especially when Ly$\alpha$ is weak or absent.
Following our previous papers \citep[e.g.,][]{p1}, we classified objects with a hint of broad spectral bump around Ly$\alpha$ as quasars, and the remaining objects as galaxies.
The line and continuum properties of the EoR objects described above (quasars, candidate obscured quasars, and galaxies) were measured as in \citet{p2}, and are listed in Table \ref{tab:specprop}.

\begin{figure*}
\epsscale{1.0}
\plotone{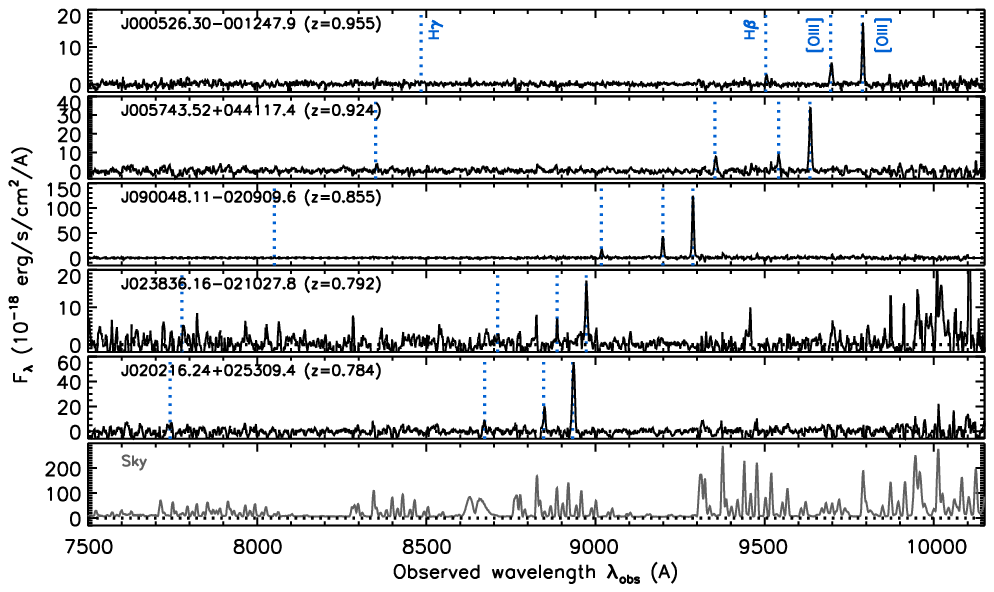}
\caption{Same as Figure \ref{fig:spec1}, but for five galaxies with strong [\ion{O}{3}] $\lambda$4959 and $\lambda$5007 lines at $0.7 < z < 1.0$.
The blue dotted lines mark the expected positions of the H$\gamma$, H$\beta$, and [\ion{O}{3}] $\lambda$4959, $\lambda$5007 emission lines, given the redshifts.
\label{fig:spec7}}
\end{figure*}

\begin{figure*}
\epsscale{1.0}
\plotone{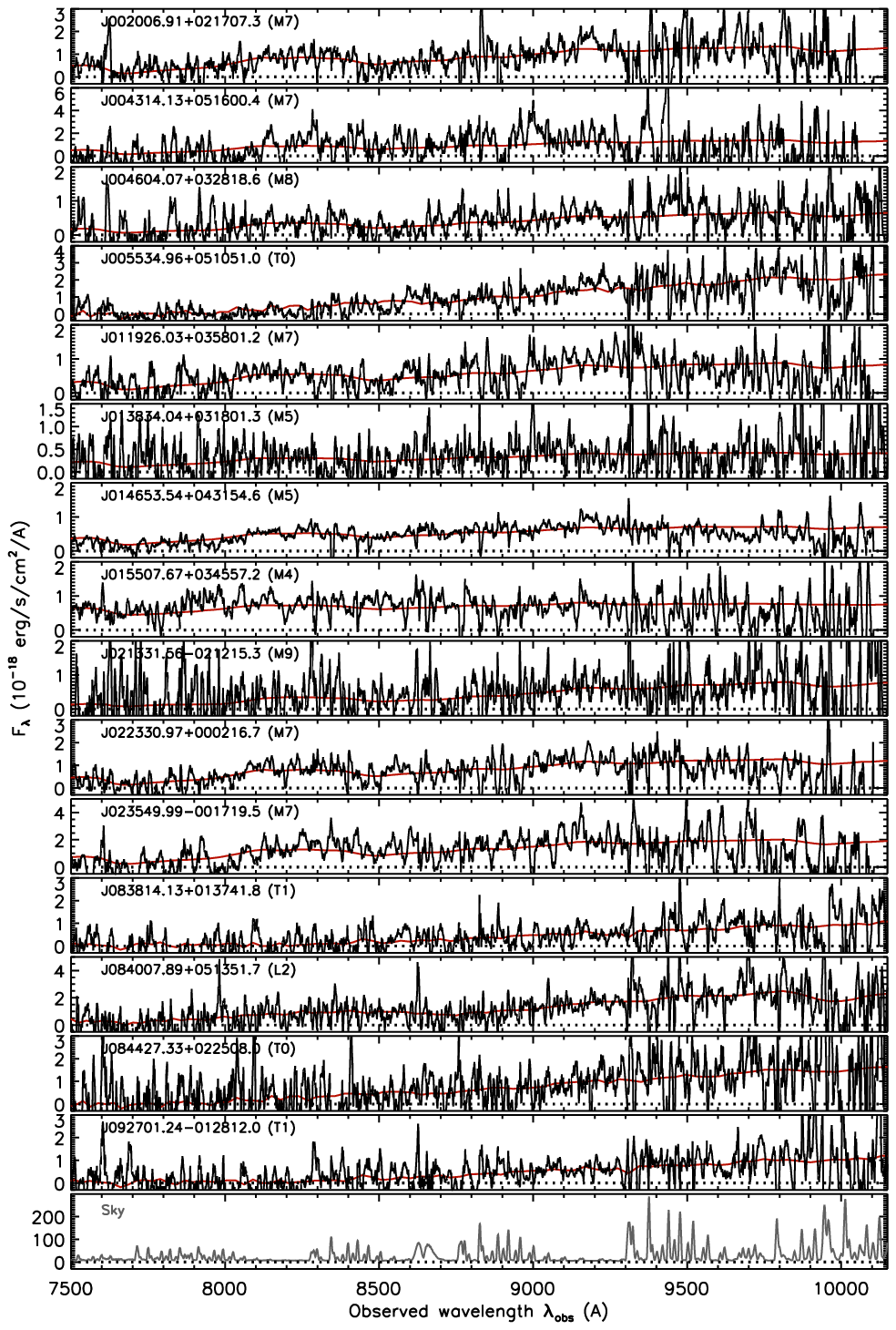}
\caption{Same as Figure \ref{fig:spec1}, but for the first set of 15 brown dwarfs, ordered by the right ascension.
The red lines represent the best-fit templates, whose spectral types are indicated at the top left corner of the individual panels.
\label{fig:spec8}}
\end{figure*}

\begin{figure*}
\epsscale{1.0}
\plotone{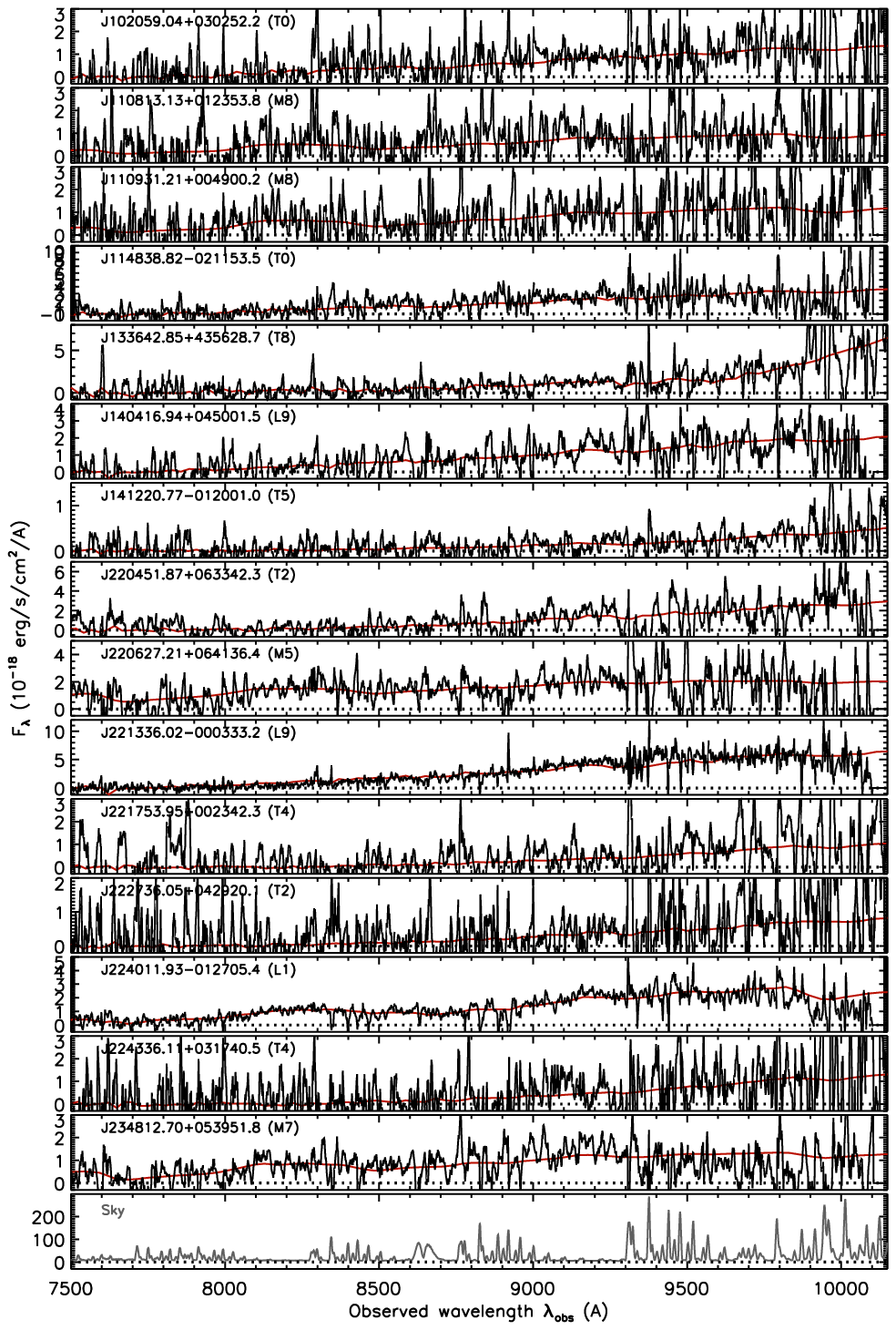}
\caption{Same as Figure \ref{fig:spec8}, but for the second set of 15 brown dwarfs.
\label{fig:spec9}}
\end{figure*}

\begin{figure*}
\epsscale{1.0}
\plotone{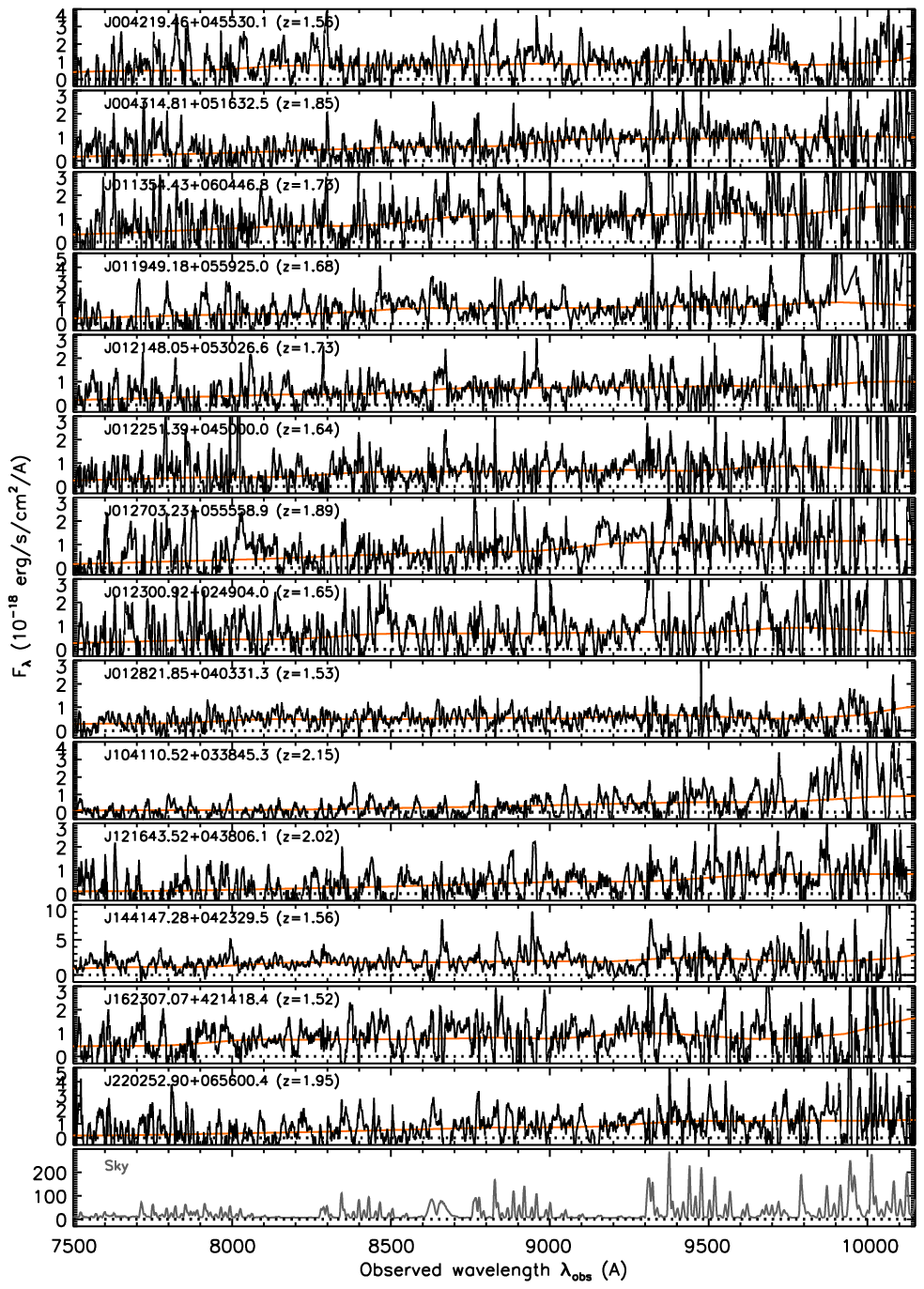}
\caption{Same as Figure \ref{fig:spec1}, but for 14 passive galaxies at $1.5 < z < 2.2$. 
The orange lines represent the best-fit templates, with the inferred redshifts indicated at the top left corner of the individual panels.
The $\chi^2$ values indicate the redshift errors up to $\Delta$z = 0.03, but the redshifts are likely more uncertain, since the fitting uses only a single template spectrum.
For more complicated modeling, we would need higher quality data over wider spectral coverage. 
\label{fig:spec10}}
\end{figure*}

Our quasar candidates also included a significant number of interlopers at lower redshifts. 
Figure \ref{fig:spec7} shows galaxies with strong  [\ion{O}{3}] $\lambda$4959 and $\lambda$5007 emission lines at $0.7 < z < 1.0$.
The [\ion{O}{3}] lines are covered by the HSC $z$ band, and the broad-band colors of these objects mimic Ly$\alpha$ of quasars at $z \sim 6$.
Their line properties are listed at the bottom of Table \ref{tab:specprop}.
Finally, we identified 44 sources with red continuum without a sharp break.
The 30 objects in Figures \ref{fig:spec8} -- \ref{fig:spec9} are point sources with our definition of $\mu$/$\mu_{\rm PSF}$ $<$ 1.2, 
and are most likely Galactic brown dwarfs.
Following our previous papers, we estimated their spectral types by fitting the spectral standard templates of M4- to T8-type dwarfs \citep{burgasser14,skrzypek15} 
to the observed spectra.
The best-fit types are indicated in the individual panels of the figure.
On the other hand, the 14 objects in Figure \ref{fig:spec10} have more extended shape with $\mu$/$\mu_{\rm PSF}$ $>$ 1.2.
It is known that passive galaxies at $z = 1 - 2$ contaminate high-$z$ quasar selection, if extended sources are not removed effectively \citep[e.g.,][]{barnett19}.
Indeed, we found that the spectra of the 14 extended objects are fitted reasonably well with a spectral template of a passive galaxy \citep{cww80} with the best-fit redshifts of $z = 1.5 - 2.2$.
The redshift estimates are very uncertain, because of the limited spectral coverage and data quality.

For the same reason,
the separation between Galactic brown dwarfs and passive galaxies remains somewhat ambiguous.
As a test, we fitted the templates of brown dwarfs and a passive galaxy to all the 44 objects with red continuum, and found that (i) a passive galaxy template is favored by only one of the
30 point sources and (ii) brown dwarf templates are favored by only three of the 14 extended sources, based on the reduced $\chi^2$ values.
This gives us some confidence that the separation between brown dwarfs and passive galaxies
is broadly correct, althouth we cannot rely too heavily on the classification of individual objects.
Near-IR photometry would be useful to clearly separate the two contaminating populations, but such an analysis is beyond the scope of this paper.

\section{Summary} \label{sec:summary}

This is the 24th publication of the SHELLQs project.
The HSC-SSP survey completed its 330 nights of observation in 2021 December, and here we show the immediate results from the follow-up spectroscopy carried out between 2021 November and 2025 March.
This paper is likely the final installment of major (unobscured) quasar discoveries from the project.
We present 132 spectra of photometric quasar candidates, which include 43 quasars at $5.71 \le z \le 7.02$, 11 candidate obscured (narrow-line) quasars at  $6.01 \le z \le 6.88$, 
and 29 galaxies without strong Ly$\alpha$ emission at similar redshifts.
The two galaxies $J140637.84+430623.2$ and $J140625.39+430504.7$ are separated by 900 kpc in projected proper distance and have identical redshifts ($z = 5.98$), 
which may indicate the presence of a galaxy overdensity in this region. 
The remaining candidates included five galaxies with strong [\ion{O}{3}] lines at $0.7 < z < 1.0$, 30 Galactic brown dwarfs, and 14 passive galaxies at $1.5 < z < 2.2$, although the separation
between the latter two classes remains somewhat ambiguous, given the quality of our spectra. 
We are currently working on a statistical analysis of the full SHELLQs sample, which will be presented in our next publication (Y. Matsuoka et al. 2025, in preparation).

\clearpage

\startlongtable
\begin{deluxetable*}{ccccccc}
\tablecaption{Spectroscopic Properties\label{tab:specprop}}
\tablehead{
\colhead{Name} & \colhead{Redshift} & 
\colhead{$M_{1450}$} & \colhead{Line} & 
\colhead{EW$_{\rm rest}$} & \colhead{FWHM} & 
\colhead{log $L_{\rm line}$}\\
\colhead{} & \colhead{} & 
\colhead{(mag)} & \colhead{} & 
\colhead{(\AA)} & \colhead{(km s$^{-1}$)} & 
\colhead{($L_{\rm line}$ in erg s$^{-1}$)}
} 
\startdata
\multicolumn{7}{c}{Quasars}\\\hline
$J133013.00+423432.2$ &  7.02 &   $-25.58 \pm 0.21$              & \nodata &               \nodata &               \nodata &               \nodata \\  
$J140344.28+423114.3$ &  6.85 &   $-24.45 \pm 0.11$           & Ly$\alpha$ &     8 $\pm$ 2 &     1700 $\pm$ 1000 &    43.56 $\pm$ 0.11 \\ 
$J225706.60+053102.4$ &  6.81 &   $-23.87 \pm 0.18$           & Ly$\alpha$ &     41 $\pm$ 9 &     940 $\pm$ 430 &    44.06 $\pm$ 0.06 \\  
$J223942.28+040943.2$ &  6.74 &   $-23.07 \pm 0.13$           & Ly$\alpha$ &     13 $\pm$ 2 &     480 $\pm$ 140 &    43.24 $\pm$ 0.03 \\  
$J015851.37+052410.3$ &  6.73 &   $-23.92 \pm 0.53$           & Ly$\alpha$ &     100 $\pm$ 60 &  11000 $\pm$ 4000 & 44.45 $\pm$ 0.12 \\  
$J142817.13+045430.3$ &  6.72 &   $-25.71 \pm 0.08$              & \nodata &               \nodata &               \nodata &               \nodata \\  
$J225810.01+035345.3$ &  6.64 &   $-26.10 \pm 0.01$           & Ly$\alpha$ &     13 $\pm$ 1 &     3000 $\pm$ 1200 &    44.43 $\pm$ 0.02 \\  
$J125934.97+021309.0$ &  6.43 &   $-24.99 \pm 0.06$           & Ly$\alpha$ &     31 $\pm$ 3 &     6300 $\pm$ 900 &    44.37 $\pm$ 0.03 \\  
$J113206.98+035011.7$ &  6.36 &   $-23.56 \pm 0.08$              & \nodata &               \nodata &               \nodata &               \nodata \\ 
$J134014.18+025855.0$ &  6.31 &   $-24.99 \pm 0.07$           & Ly$\alpha$ &     17 $\pm$ 3 &     1100 $\pm$ 1300 &    44.11 $\pm$ 0.06 \\  
$J221858.14+062406.3$ &  6.29 &   $-24.31 \pm 0.11$           & Ly$\alpha$ &     20 $\pm$ 3 &     1900 $\pm$ 600 &    43.92 $\pm$ 0.05 \\ 
$J132905.06+431640.4$ &  6.29 &   $-23.87 \pm 0.10$           & Ly$\alpha$ &     95 $\pm$ 11 &     2000 $\pm$ 2100 &    44.40 $\pm$ 0.03 \\ 
$J133534.37+025702.7$ &  6.27 &   $-24.20 \pm 0.07$           & Ly$\alpha$ &     15 $\pm$ 2 &     2300 $\pm$ 300 &    43.76 $\pm$ 0.05 \\ 
$J113007.51+051149.1$ &  6.27 &   $-25.16 \pm 0.02$              & \nodata &               \nodata &               \nodata &               \nodata \\  
$J094624.10-001233.7$ &  6.26 &   $-22.99 \pm 0.14$           & Ly$\alpha$ &     21 $\pm$ 4 &          440 $\pm$ 630 &    43.43 $\pm$ 0.05 \\ 
$J111123.42+034039.4$ &  6.26 &   $-22.46 \pm 0.19$           & Ly$\alpha$ &     260 $\pm$ 50 &     660 $\pm$ 110 &    44.30 $\pm$ 0.01 \\ 
$J013603.22+022605.0$ &  6.21 &   $-24.46 \pm 0.15$           & Ly$\alpha$ &     95 $\pm$ 14 &     670 $\pm$ 390 &    44.66 $\pm$ 0.02 \\  
$J023558.20+043456.8$ &  6.18 &   $-23.65 \pm 0.10$           & Ly$\alpha$ &     41 $\pm$ 8 &     5000 $\pm$ 3300 &    43.95 $\pm$ 0.07 \\ 
$J125139.16+023342.2$ &  6.17 &   $-21.53 \pm 0.94$           & Ly$\alpha$ &     780 $\pm$ 670 &     780 $\pm$ 220 &    44.41 $\pm$ 0.01 \\ 
$J135818.89-013249.7$ &  6.14 &   $-22.18 \pm 0.64$           & Ly$\alpha$ &     200 $\pm$ 120 &     430 $\pm$ 130 &    44.07 $\pm$ 0.02 \\  
$J133740.18+434557.6$ &  6.14 &   $-23.11 \pm 0.05$           & Ly$\alpha$ &     6 $\pm$ 1         &     710 $\pm$ 200 &    42.93 $\pm$ 0.04 \\ 
$J141305.38+035829.9$ &  6.12 &   $-23.46 \pm 0.11$              & \nodata &               \nodata &               \nodata &               \nodata \\ 
$J105642.67+041902.1$ &  6.12 &   $-21.59 \pm 0.31$           & Ly$\alpha$ &     270 $\pm$ 80 &     550 $\pm$ 380 &    43.98 $\pm$ 0.01 \\ 
$J020038.83-021052.0$ &  6.11 &   $-21.43 \pm 0.30$           & Ly$\alpha$ &     420 $\pm$ 120 &     700 $\pm$ 130 &    44.08 $\pm$ 0.01 \\ 
$J022215.99-012534.4$ &  6.11 &   $-22.66 \pm 0.13$           & Ly$\alpha$ &     29 $\pm$ 5 &     1200 $\pm$ 1100 &    43.42 $\pm$ 0.06 \\   
$J020236.24+030033.6$ &  6.10 &   $-23.29 \pm 0.09$           & Ly$\alpha$ &     10 $\pm$ 2 &     800 $\pm$ 480 &    43.21 $\pm$ 0.08 \\   
$J131609.75+003553.6$ &  6.09 &   $-23.51 \pm 0.14$           & Ly$\alpha$ &     9 $\pm$ 3 &     1300 $\pm$ 700 &    43.24 $\pm$ 0.13 \\  
$J134021.30+031939.2$ &  6.06 &   $-22.87 \pm 0.11$           & Ly$\alpha$ &     64 $\pm$ 7 &     3400 $\pm$ 1600 &    43.85 $\pm$ 0.02 \\ 
$J001705.78+004343.3$ &  6.05 &   $-21.91 \pm 0.32$           & Ly$\alpha$ &     410 $\pm$ 120 &     790 $\pm$ 90 &    44.25 $\pm$ 0.01 \\ 
$J015812.02+002436.1$ &  6.05 &   $-23.33 \pm 0.07$           & Ly$\alpha$ &     61 $\pm$ 5 &     4200 $\pm$ 1700 &    44.01 $\pm$ 0.02 \\   
$J115900.87+025819.2$ &  6.05 &   $-22.82 \pm 0.08$           & Ly$\alpha$ &     600 $\pm$ 530 &     430 $\pm$ 30 &    44.17 $\pm$ 0.01 \\  
$J220754.87+065111.8$ &  6.05 &   $-24.44 \pm 0.08$           & Ly$\alpha$ &     84 $\pm$ 7 &     1100 $\pm$ 100 &    44.60 $\pm$ 0.01 \\  
$J220754.87+065111.8$ &   \nodata  & \nodata                        & NV             &     33 $\pm$ 4 &     1800 $\pm$ 600 &    44.20 $\pm$ 0.03 \\  
$J011830.06+023642.1$ &  6.01 &   $-23.70 \pm 0.04$           & Ly$\alpha$ &     32 $\pm$ 3 &     6100 $\pm$ 1700 &    43.88 $\pm$ 0.03 \\   
$J124642.21+051048.9$ &  6.01 &   $-24.37 \pm 0.06$           & Ly$\alpha$ &     61 $\pm$ 5 &     7800 $\pm$ 900 &    44.43 $\pm$ 0.02 \\  
$J003658.18+034313.2$ &  6.00 &   $-23.29 \pm 0.12$           & Ly$\alpha$ &     100 $\pm$ 10 &     2400 $\pm$ 200 &    44.22 $\pm$ 0.02 \\ 
$J131830.28+051231.7$ &  6.00 &   $-24.26 \pm 0.05$           & Ly$\alpha$ &     16 $\pm$ 2 &     4700 $\pm$ 600 &    43.82 $\pm$ 0.04 \\  
$J142421.11+040512.1$ &  5.99 &   $-24.53 \pm 0.03$           & Ly$\alpha$ &     56 $\pm$ 3 &     10000 $\pm$ 1000 &    44.45 $\pm$ 0.02 \\  
$J023811.54+040416.5$ &  5.98 &   $-24.67 \pm 0.09$           & Ly$\alpha$ &     6 $\pm$ 1 &     300 $\pm$ 300 &           43.57 $\pm$ 0.05 \\  
$J003604.19+040912.0$ &  5.98 &   $-23.05 \pm 0.07$           & Ly$\alpha$ &     88 $\pm$ 7 &     750 $\pm$ 110 &    44.05 $\pm$ 0.02 \\ 
$J012031.64+004226.7$ &  5.96 &   $-23.08 \pm 0.08$           & Ly$\alpha$ &     53 $\pm$ 6 &     2900 $\pm$ 1100 &    43.84 $\pm$ 0.04 \\ 
$J130518.55+030437.5$ &  5.93 &   $-24.52 \pm 0.05$              & \nodata &               \nodata &               \nodata &               \nodata \\ 
$J133611.60-011929.6$ &  5.92 &   $-23.94 \pm 0.03$           & Ly$\alpha$ &     0.4 $\pm$ 0.4 &     560 $\pm$ 290 &    42.07 $\pm$ 0.39 \\  
$J224357.80+061816.1$ &  5.71 &   $-23.46 \pm 0.07$           & Ly$\alpha$ &     5 $\pm$ 1 &          650 $\pm$ 170 &    42.99 $\pm$ 0.06 \\  
\hline\multicolumn{7}{c}{Candidate Obscured Quasars}\\\hline
$J142322.01+020612.8$ &  6.88 &   $-24.28 \pm 0.34$           & Ly$\alpha$ &     67 $\pm$ 21 &             $<$ 230 &    44.33 $\pm$ 0.02 \\  
$J125001.43+015243.2$ &  6.29 &   $-21.67 \pm 0.62$           & Ly$\alpha$ &     270 $\pm$ 150 &     300 $\pm$ 100 & 43.89 $\pm$ 0.02 \\ 
$J131304.72+003542.2$ &  6.23 &   $-22.60 \pm 0.12$           & Ly$\alpha$ &     120 $\pm$ 10 &     270 $\pm$ 10 &    43.92 $\pm$ 0.01 \\  
$J125456.03-005857.9$ &  6.17 &   $-23.30 \pm 0.07$           & Ly$\alpha$ &     11 $\pm$ 1 &             $<$ 230 &        43.14 $\pm$ 0.03 \\  
$J020719.59+023826.0$ &  6.14 &   $-20.23 \pm 1.59$           & Ly$\alpha$ &     840 $\pm$ 830 &     400 $\pm$ 30 &    44.06 $\pm$ 0.01 \\ 
$J022336.97-015636.1$ &  6.13 &   $-23.00 \pm 0.08$           & Ly$\alpha$ &      51 $\pm$ 4 &     450 $\pm$ 70 &    43.70 $\pm$ 0.01 \\ 
$J132535.14+012947.2$ &  6.10 &   $-23.44 \pm 0.07$           & Ly$\alpha$ &     15 $\pm$ 2 &     330 $\pm$ 140 &    43.34 $\pm$ 0.05 \\  
$J133846.93-005339.5$ &  6.09 &   $-22.69 \pm 0.20$           & Ly$\alpha$ &     88 $\pm$ 17 &     330 $\pm$ 150 &    43.81 $\pm$ 0.02 \\  
$J131448.69+031106.4$ &  6.04 &   $-22.92 \pm 0.09$           & Ly$\alpha$ &     36 $\pm$ 3 &     270 $\pm$ 30 &    43.51 $\pm$ 0.01 \\  
$J143905.32+032356.0$ &  6.03 &   $-22.77 \pm 0.09$           & Ly$\alpha$ &     66 $\pm$ 6 &     240 $\pm$ 30 &    43.72 $\pm$ 0.01 \\  
$J015158.79+043325.6$ &  6.01 &   $-23.01 \pm 0.16$           & Ly$\alpha$ &     16 $\pm$ 3 &     390 $\pm$ 60 &    43.27 $\pm$ 0.03 \\ 
\hline\multicolumn{7}{c}{Galaxies}\\\hline
$J013633.12+003410.3$ &  6.28 &   $-23.29 \pm 0.08$              & \nodata &               \nodata &               \nodata &               \nodata \\ 
$J100702.44+050545.9$ &  6.25 &   $-23.82 \pm 0.06$              & \nodata &               \nodata &               \nodata &               \nodata \\  
$J013958.57+034539.5$ &  6.10 &   $-23.25 \pm 0.07$              & \nodata &               \nodata &               \nodata &               \nodata \\ 
$J135933.30+025909.3$ &  6.09 &   $-23.00 \pm 0.11$           & Ly$\alpha$ &          7 $\pm$ 1 &     390 $\pm$ 310 &    42.85 $\pm$ 0.08 \\ 
$J233252.97+024353.0$ &  6.08 &   $-23.57 \pm 0.11$              & \nodata &               \nodata &               \nodata &               \nodata \\   
$J133605.10+022147.1$ &  6.02 &   $-23.23 \pm 0.07$              & \nodata &               \nodata &               \nodata &               \nodata \\ 
$J112550.16-001619.6$ &  6.01 &   $-24.25 \pm 0.30$              & \nodata &               \nodata &               \nodata &               \nodata \\  
$J140637.84+430623.2$  &  5.98 &   $-23.25 \pm 0.05$  & Ly$\alpha$ & 1.3 $\pm$ 0.4 & 220 $\pm$ 210 &    42.27 $\pm$ 0.12 \\ 
$J140625.39+430504.7$  &  5.98 &   $-22.28 \pm 0.09$  & Ly$\alpha$ &     6 $\pm$ 1   & 480 $\pm$ 160 &    42.54 $\pm$ 0.08 \\ 
$J020213.75+011910.1$ &  5.94 &   $-23.01 \pm 0.18$              & \nodata &               \nodata &               \nodata &               \nodata \\ 
$J084902.00-013025.7$ &  5.94 &   $-23.00 \pm 0.12$              & \nodata &               \nodata &               \nodata &               \nodata \\ 
$J235527.98+024447.4$ &  5.93 &   $-23.38 \pm 0.14$              & \nodata &               \nodata &               \nodata &               \nodata \\ 
$J132746.13-015410.3$ &  5.92 &   $-23.38 \pm 0.08$              & \nodata &               \nodata &               \nodata &               \nodata \\  
$J132514.41+034006.0$ &  5.92 &   $-23.25 \pm 0.06$           & Ly$\alpha$ &     0.8 $\pm$ 0.3 &             $<$ 230 &    42.05 $\pm$ 0.16 \\  
$J104800.36+011856.4$ &  5.91 &   $-23.12 \pm 0.15$              & \nodata &               \nodata &               \nodata &               \nodata \\ 
$J021103.93+015516.9$ &  5.91 &   $-23.40 \pm 0.13$              & \nodata &               \nodata &               \nodata &               \nodata \\   
$J135059.71+023717.3$ &  5.90 &   $-22.74 \pm 0.10$              & \nodata &               \nodata &               \nodata &               \nodata \\  
$J014056.95-005048.7$ &  5.89 &   $-23.27 \pm 0.07$              & \nodata &               \nodata &               \nodata &               \nodata \\   
$J131024.32+005002.0$ &  5.88 &   $-23.16 \pm 0.08$              & \nodata &               \nodata &               \nodata &               \nodata \\ 
$J003554.15+030807.3$ &  5.87 &   $-24.27 \pm 0.04$              & \nodata &               \nodata &               \nodata &               \nodata \\ 
$J120542.50+041430.1$ &  5.86 &   $-23.14 \pm 0.21$              & \nodata &               \nodata &               \nodata &               \nodata \\ 
$J103825.94-004418.9$ &  5.86 &   $-23.22 \pm 0.06$              & \nodata &               \nodata &               \nodata &               \nodata \\  
$J124359.28+015305.5$ &  5.86 &   $-23.07 \pm 0.07$              & \nodata &               \nodata &               \nodata &               \nodata \\  
$J000928.93+001644.7$ &  5.83 &   $-23.16 \pm 0.10$              & \nodata &               \nodata &               \nodata &               \nodata \\ 
$J234434.18+022645.7$ &  5.83 &   $-23.50 \pm 0.09$              & \nodata &               \nodata &               \nodata &               \nodata \\ 
$J123934.04+025845.8$ &  5.82 &   $-22.92 \pm 0.08$           & Ly$\alpha$ &          4 $\pm$ 1 &     450 $\pm$ 70   &    42.68 $\pm$ 0.11 \\  
$J000440.93+010500.8$ &  5.81 &   $-23.06 \pm 0.04$              & \nodata &               \nodata &               \nodata &               \nodata \\ 
$J113411.21-001956.1$ &  5.74 &   $-23.42 \pm 0.11$              & \nodata &               \nodata &               \nodata &               \nodata \\ 
$J145945.34+415206.4$ &  5.56 &   $-24.62 \pm 0.13$              & \nodata &               \nodata &               \nodata &               \nodata \\  
\hline\multicolumn{7}{c}{[\ion{O}{3}] Emitters}\\\hline
$J000526.30-001247.9$ & 0.955 &  \nodata    & H$\gamma$   &               \nodata &               \nodata &               \nodata \\  
                      &       &               \nodata             & H$\beta$        &     150 $\pm$ 40 &             $<$ 230 &    40.90 $\pm$ 0.08 \\  
                      &       &               \nodata & [OIII] $\lambda$4959 &     450 $\pm$ 100 &             $<$ 230 &    41.37 $\pm$ 0.03 \\  
                      &       &               \nodata & [OIII] $\lambda$5007 &     1000 $\pm$ 200 &             $<$ 230 &    41.73 $\pm$ 0.02 \\  
$J005743.52+044117.4$ & 0.924 &  \nodata     & H$\gamma$ &     170 $\pm$ 90 &             $<$ 230 &    41.07 $\pm$ 0.09 \\ 
                      &       &               \nodata             & H$\beta$        &     390 $\pm$ 190 &             $<$ 230 &    41.43 $\pm$ 0.03 \\ 
                      &       &               \nodata & [OIII] $\lambda$4959 &     340 $\pm$ 160 &             $<$ 230 &    41.37 $\pm$ 0.03 \\ 
                      &       &               \nodata & [OIII] $\lambda$5007 &     2000 $\pm$ 900 &             $<$ 230 &    42.14 $\pm$ 0.01 \\ 
$J090048.11-020909.6$ & 0.855 &  \nodata     & H$\gamma$ &               \nodata &               \nodata &               \nodata \\ 
                      &       &               \nodata             & H$\beta$       &     770 $\pm$ 380 &             $<$ 190 &    41.62 $\pm$ 0.02 \\ 
                      &       &               \nodata & [OIII] $\lambda$4959 &     1900 $\pm$ 900 &             $<$ 190 &    42.01 $\pm$ 0.01 \\ 
                      &       &               \nodata & [OIII] $\lambda$5007 &     5900 $\pm$ 2900 &             $<$ 190 &    42.50 $\pm$ 0.01 \\ 
$J023836.16-021027.8$ & 0.792 &               \nodata            & H$\gamma$ &               \nodata &               \nodata &               \nodata \\ 
                      &       &               \nodata             & H$\beta$ &               \nodata &               \nodata &               \nodata \\ 
                      &       &               \nodata & [OIII] $\lambda$4959 &               $>$ 90 &             $<$ 230 &    41.177 $\pm$ 0.173 \\ 
                      &       &               \nodata & [OIII] $\lambda$5007 &               $>$ 240 &             $<$ 230 &    41.582 $\pm$ 0.020 \\ 
$J020216.24+025309.4$ & 0.784 &               \nodata            & H$\gamma$ &               \nodata &               \nodata &               \nodata \\ 
                      &       &               \nodata             & H$\beta$ &                      $>$ 70 &             210 $\pm$ 60 &    41.216 $\pm$ 0.098 \\ 
                      &       &               \nodata & [OIII] $\lambda$4959 &               $>$ 240 &             $<$ 230 &    41.735 $\pm$ 0.039 \\ 
                      &       &               \nodata & [OIII] $\lambda$5007 &               $>$ 620 &             $<$ 230 &    42.141 $\pm$ 0.015 \\ 
\enddata
\tablecomments{
The redshifts of quasars, candidate obscured quasars, and galaxies have uncertainties $\Delta z \sim 0.01 - 0.1$, depending on the spectral features around Ly$\alpha$
\citep[e.g.,][]{p16}.
``EW$_{\rm rest}$" represents the rest-frame equivalent width; 3$\sigma$ lower limits are reported for objects without detectable continuum.}
\end{deluxetable*}

\clearpage

\acknowledgments

This research is based on data collected at Subaru Telescope, which is operated by the National Astronomical Observatory of Japan. 
We are honored and grateful for the opportunity of observing the Universe from Maunakea, which has the cultural, historical and natural significance in Hawaii.
We appreciate the staff members of the telescope for their support during our FOCAS observations.
The data analysis was in part carried out on the open use data analysis computer system at the Astronomy Data Center of NAOJ.

This work is also based on observations made with the GTC, installed at the Spanish Observatorio del Roque de los Muchachos 
of the Instituto de Astrof\'{i}sica de Canarias, on the island of La Palma.
We thank the support astronomers for their help during preparation and execution of our observing program.

Y. M. was supported by the Japan Society for the Promotion of Science (JSPS) KAKENHI Grant No. 21H04494. 
K. I. acknowledges the support under the grant PID2022-136827NB-C44 provided by MCIN/AEI /10.13039/501100011033 / FEDER, EU.
M. O was supported by the JSPS KAKENHI Grant No. G24K22894.

The HSC collaboration includes the astronomical communities of Japan and Taiwan, and Princeton University.  The HSC instrumentation and software were developed by the National Astronomical Observatory of Japan (NAOJ), the Kavli Institute for the Physics and Mathematics of the Universe (Kavli IPMU), the University of Tokyo, the High Energy Accelerator Research Organization (KEK), the Academia Sinica Institute for Astronomy and Astrophysics in Taiwan (ASIAA), and Princeton University.  Funding was contributed by the FIRST program from the Japanese Cabinet Office, the Ministry of Education, Culture, Sports, Science and Technology (MEXT), the Japan Society for the Promotion of Science (JSPS), Japan Science and Technology Agency  (JST), the Toray Science  Foundation, NAOJ, Kavli IPMU, KEK, ASIAA, and Princeton University.
 
This paper is based on data retrieved from the HSC data archive system, which is operated by Subaru Telescope and Astronomy Data Center (ADC) at NAOJ. 
Data analysis was in part carried out with the cooperation of Center for Computational Astrophysics (CfCA) at NAOJ.   

This paper makes use of software developed for Vera C. Rubin Observatory. We thank the Rubin Observatory for making their code available as free software at http://pipelines.lsst.io/. 
 
The Pan-STARRS1 Surveys (PS1) and the PS1 public science archive have been made possible through contributions by the Institute for Astronomy, the University of Hawaii, the Pan-STARRS Project Office, the Max Planck Society and its participating institutes, the Max Planck Institute for Astronomy, Heidelberg, and the Max Planck Institute for Extraterrestrial Physics, Garching, The Johns Hopkins University, Durham University, the University of Edinburgh, the Queen's University Belfast, the Harvard-Smithsonian Center for Astrophysics, the Las Cumbres Observatory Global Telescope Network Incorporated, the National Central University of Taiwan, the Space Telescope Science Institute, the National Aeronautics and Space Administration under grant No. NNX08AR22G issued through the Planetary Science Division of the NASA Science Mission Directorate, the National Science Foundation grant No. AST-1238877, the University of Maryland, Eotvos Lorand University (ELTE), the Los Alamos National Laboratory, and the Gordon and Betty Moore Foundation.

\appendix

We present the journal of spectroscopic observations in Table \ref{tab:obsjournal}.

\startlongtable
\begin{deluxetable*}{rcccrl}
\tablecaption{Journal of Discovery Spectroscopy \label{tab:obsjournal}}
\tablehead{
\colhead{Object} & 
\colhead{$i_{\rm AB}$} & \colhead{$z_{\rm AB}$} & \colhead{$y_{\rm AB}$} & \colhead{$t_{\rm exp}$} & \colhead{Date (Inst)}\\
\colhead{} & 
\colhead{(mag)} & \colhead{(mag)} & \colhead{(mag)} & \colhead{(min)} & \colhead{}
} 
\startdata
\multicolumn{6}{c}{Quasars}\\\hline
$J133013.00+423432.2$  &  $>$26.10   &  $>$25.05   &  22.67 $\pm$  0.05  &   60  &     2025 Jan 28 (O)\\
$J140344.28+423114.3$  &  $>$26.06   &  $>$25.04   &  22.61 $\pm$  0.05  &   30  &     2022 Apr 24 (F)\\
$J225706.60+053102.4$  &  $>$25.68   &  $>$25.00   &  22.72 $\pm$  0.07  &   60  &     2024 Sep 25 (O)\\
$J223942.28+040943.2$  &  $>$26.14   &  $>$25.58   &  23.57 $\pm$  0.09  &  120  &     2024 Nov 06 (F)\\
$J015851.37+052410.3$  &  $>$25.04   &  25.37 $\pm$  0.79  &  22.17 $\pm$  0.07  &   60  &     2024 Oct 29 (O)\\
$J142817.13+045430.3$  &  25.81 $\pm$  0.32  &  $>$24.24   &  21.26 $\pm$  0.02  &   60  &     2025 Jan 24 (O)\\
$J225810.01+035345.3$  &  25.51 $\pm$  0.16  &  23.38 $\pm$  0.08  &  20.80 $\pm$  0.01  &   20  &     2021 Nov 25 (F)\\
$J125934.97+021309.0$  &  25.52 $\pm$  0.11  &  22.42 $\pm$  0.03  &  21.75 $\pm$  0.02  &   15  &     2023 Jan 10 (F)\\
$J113206.98+035011.7$  &  27.47 $\pm$  0.74  &  23.84 $\pm$  0.07  &  23.44 $\pm$  0.08  &   40  &     2022 Dec 13 (F)\\
$J134014.18+025855.0$  &  25.37 $\pm$  0.13  &  22.22 $\pm$  0.02  &  21.80 $\pm$  0.02  &   20  &     2023 Jan 11 (F)\\
$J221858.14+062406.3$  &  25.64 $\pm$  0.16  &  22.44 $\pm$  0.04  &  22.44 $\pm$  0.05  &   20  &     2024 Aug 10 (O)\\
$J132905.06+431640.4$  &  26.76 $\pm$  0.35  &  22.65 $\pm$  0.02  &  23.17 $\pm$  0.08  &   20  &     2022 Dec 27 (F)\\
$J133534.37+025702.7$  &  25.87 $\pm$  0.28  &  22.86 $\pm$  0.03  &  22.94 $\pm$  0.05  &   15  &     2023 Jan 10 (F)\\
$J113007.51+051149.1$  &  25.51 $\pm$  0.18  &  22.22 $\pm$  0.02  &  21.59 $\pm$  0.02  &   15  &     2022 Dec 13 (F)\\
$J094624.10-001233.7$  &  $>$26.27   &  24.01 $\pm$  0.07  &  23.87 $\pm$  0.14  &   30  &     2022 Dec 13 (F)\\
$J111123.42+034039.4$  &  $>$26.50   &  23.13 $\pm$  0.02  &  24.90 $\pm$  0.30  &   60  &     2023 Mar 17 (O)\\
$J013603.22+022605.0$  &  25.19 $\pm$  0.11  &  21.61 $\pm$  0.01  &  21.98 $\pm$  0.02  &   15  &     2024 Aug 12 (O)\\
$J023558.20+043456.8$  &  25.97 $\pm$  0.32  &  23.01 $\pm$  0.04  &  22.99 $\pm$  0.11  &   45  &     2024 Sep 13 (O)\\
$J125139.16+023342.2$  &  $>$25.88   &  22.85 $\pm$  0.03  &  24.54 $\pm$  0.19  &   10  &     2022 Dec 13 (F)\\
$J135818.89-013249.7$  &  26.87 $\pm$  0.38  &  23.66 $\pm$  0.05  &  24.10 $\pm$  0.20  &   60  &     2024 Jul 01 (O)\\
$J133740.18+434557.6$  &  26.41 $\pm$  0.30  &  23.54 $\pm$  0.06  &  23.34 $\pm$  0.09  &   30  &     2022 Dec 26 (F)\\
$J141305.38+035829.9$  &  25.88 $\pm$  0.22  &  23.50 $\pm$  0.06  &  23.58 $\pm$  0.12  &   67  &     2025 Feb 04 (O)\\
$J105642.67+041902.1$  &  $>$26.25   &  23.85 $\pm$  0.05  &  25.51 $\pm$  0.68  &   30  &     2024 Nov 08 (F)\\
$J020038.83-021052.0$  &  $>$26.04   &  23.67 $\pm$  0.05  &  26.49 $\pm$  0.99  &   20  &     2021 Nov 25 (F)\\
$J022215.99-012534.4$  &  $>$25.87   &  23.98 $\pm$  0.06  &  24.06 $\pm$  0.13  &   50  &     2021 Dec 25 (F)\\
$J020236.24+030033.6$  &  27.11 $\pm$  0.59  &  23.87 $\pm$  0.06  &  23.89 $\pm$  0.19  &   30  &     2021 Dec 25 (F)\\
$J131609.75+003553.6$  &  26.45 $\pm$  0.34  &  23.40 $\pm$  0.04  &  23.28 $\pm$  0.08  &   60  &     2023 Mar 17 (O)\\
$J134021.30+031939.2$  &  26.21 $\pm$  0.25  &  23.26 $\pm$  0.04  &  23.74 $\pm$  0.10  &   30  &     2023 Jan 10 (F)\\
$J001705.78+004343.3$  &  26.08 $\pm$  0.19  &  23.35 $\pm$  0.04  &  24.88 $\pm$  0.40  &   10  &     2021 Nov 25 (F)\\
$J015812.02+002436.1$  &  25.36 $\pm$  0.11  &  22.92 $\pm$  0.03  &  23.46 $\pm$  0.08  &   15  &     2021 Nov 27 (F)\\
$J115900.87+025819.2$  &  26.53 $\pm$  0.39  &  23.13 $\pm$  0.04  &  24.32 $\pm$  0.15  &   60  &     2023 Mar 15 (O)\\
$J220754.87+065111.8$  &  24.29 $\pm$  0.04  &  21.59 $\pm$  0.01  &  22.24 $\pm$  0.05  &   30  &     2024 Nov 06 (F)\\
$J011830.06+023642.1$  &  25.34 $\pm$  0.13  &  22.85 $\pm$  0.03  &  22.81 $\pm$  0.06  &   15  &     2021 Nov 27 (F)\\
$J124642.21+051048.9$  &  24.53 $\pm$  0.07  &  21.94 $\pm$  0.02  &  22.09 $\pm$  0.05  &   15  &     2022 Dec 27 (F)\\
$J003658.18+034313.2$  &  24.90 $\pm$  0.08  &  22.86 $\pm$  0.02  &  23.03 $\pm$  0.05  &   25  &     2021 Nov 27 (F)\\
$J131830.28+051231.7$  &  24.55 $\pm$  0.07  &  22.32 $\pm$  0.03  &  22.16 $\pm$  0.04  &   25  &     2023 Jan 10 (F)\\
$J142421.11+040512.1$  &  24.16 $\pm$  0.04  &  21.85 $\pm$  0.01  &  21.93 $\pm$  0.02  &   15  &     2022 May 26 (O)\\
$J023811.54+040416.5$  &  24.72 $\pm$  0.18  &  22.00 $\pm$  0.02  &  22.08 $\pm$  0.05  &   15  &     2024 Sep 13 (O)\\
$J003604.19+040912.0$  &  24.94 $\pm$  0.08  &  23.32 $\pm$  0.04  &  23.70 $\pm$  0.10  &   20  &     2021 Nov 25 (F)\\
$J012031.64+004226.7$  &  25.45 $\pm$  0.12  &  23.45 $\pm$  0.05  &  23.59 $\pm$  0.08  &   20  &     2021 Nov 25 (F)\\
$J130518.55+030437.5$  &  23.74 $\pm$  0.02  &  21.96 $\pm$  0.01  &  21.97 $\pm$  0.02  &   30  &     2023 Jan 11 (F)\\
$J133611.60-011929.6$  &  24.70 $\pm$  0.06  &  22.70 $\pm$  0.02  &  22.25 $\pm$  0.02  &   30  &     2024 Mar 29 (F)\\
$J224357.80+061816.1$  &  24.79 $\pm$  0.15  &  23.25 $\pm$  0.06  &  23.48 $\pm$  0.15  &   50  &     2024 Nov 08 (F)\\
\hline\multicolumn{6}{c}{Candidate Obscured Quasars}\\\hline
$J142322.01+020612.8$  &  $>$26.09   &  $>$25.45   &  23.63 $\pm$  0.08  &   70  &     2022 Apr 23 (F)\\
$J125001.43+015243.2$  &  $>$25.92   &  23.87 $\pm$  0.05  &  24.86 $\pm$  0.25  &   60  &     2023 Mar 17 (O)\\
$J131304.72+003542.2$  &  $>$26.07   &  23.83 $\pm$  0.07  &  24.37 $\pm$  0.22  &   30  &     2024 Mar 29 (F)\\
$J125456.03-005857.9$  &  26.93 $\pm$  0.49  &  23.92 $\pm$  0.08  &  24.02 $\pm$  0.19  &   30  &     2024 Mar 29 (F)\\
$J020719.59+023826.0$  &  $>$26.00   &  23.67 $\pm$  0.05  &  24.95 $\pm$  0.61  &   15  &     2021 Nov 26 (F)\\
$J022336.97-015636.1$  &  26.02 $\pm$  0.19  &  23.70 $\pm$  0.04  &  23.93 $\pm$  0.11  &   30  &     2021 Dec 24 (F)\\
$J132535.14+012947.2$  &  26.02 $\pm$  0.22  &  23.59 $\pm$  0.04  &  23.60 $\pm$  0.08  &   50  &     2024 Mar 28 (F)\\
$J133846.93-005339.5$  &  $>$26.15   &  23.72 $\pm$  0.05  &  23.65 $\pm$  0.09  &   30  &     2024 Mar 28 (F)\\
$J131448.69+031106.4$  &  26.82 $\pm$  0.31  &  23.78 $\pm$  0.05  &  23.78 $\pm$  0.10  &   30  &     2024 Mar 29 (F)\\
$J143905.32+032356.0$  &  26.08 $\pm$  0.23  &  23.73 $\pm$  0.06  &  24.21 $\pm$  0.25  &   30  &     2024 Mar 30 (F)\\
$J015158.79+043325.6$  &  25.82 $\pm$  0.19  &  23.81 $\pm$  0.07  &  24.15 $\pm$  0.22  &   30  &     2024 Nov 06 (F)\\
\hline\multicolumn{6}{c}{Galaxies}\\\hline
$J013633.12+003410.3$  &  $>$26.03   &  23.82 $\pm$  0.07  &  24.17 $\pm$  0.16  &   25  &     2021 Nov 26 (F)\\
$J100702.44+050545.9$  &  27.22 $\pm$  0.92  &  23.88 $\pm$  0.11  &  23.31 $\pm$  0.08  &   30  &     2024 Mar 30 (F)\\
$J013958.57+034539.5$  &  27.41 $\pm$  0.76  &  23.86 $\pm$  0.07  &  23.48 $\pm$  0.11  &   50  &     2024 Nov 07 (F)\\
$J135933.30+025909.3$  &  27.52 $\pm$  0.75  &  23.93 $\pm$  0.07  &  23.71 $\pm$  0.13  &   50  &     2023 Jan 10 (F)\\
$J233252.97+024353.0$  &  27.52 $\pm$  0.61  &  23.67 $\pm$  0.07  &  23.49 $\pm$  0.08  &   30  &     2021 Dec 25 (F)\\
$J133605.10+022147.1$  &  25.32 $\pm$  0.10  &  23.21 $\pm$  0.04  &  23.19 $\pm$  0.06  &   30  &     2022 Dec 27 (F)\\
$J112550.16-001619.6$  &  26.33 $\pm$  0.22  &  23.89 $\pm$  0.07  &  24.11 $\pm$  0.20  &   80  &     2025 Mar 21 (O)\\
$J140637.84+430623.2$  &  25.70 $\pm$  0.16  &  23.68 $\pm$  0.06  &  23.46 $\pm$  0.11  &   30  &     2022 Dec 26 (F)\\
$J140625.39+430504.7$  &  26.65 $\pm$  0.36  &  24.34 $\pm$  0.09  &  24.16 $\pm$  0.20  &   30  &     2022 Dec 26 (F)\\
$J020213.75+011910.1$  &  25.51 $\pm$  0.14  &  23.76 $\pm$  0.07  &  24.33 $\pm$  0.18  &   30  &     2021 Dec 25 (F)\\
$J084902.00-013025.7$  &  25.29 $\pm$  0.10  &  23.56 $\pm$  0.04  &  23.35 $\pm$  0.07  &   50  &     2022 Dec 27 (F)\\
$J235527.98+024447.4$  &  25.62 $\pm$  0.14  &  23.76 $\pm$  0.06  &  23.98 $\pm$  0.14  &   30  &     2021 Dec 25 (F)\\
$J132746.13-015410.3$  &  25.79 $\pm$  0.28  &  23.54 $\pm$  0.07  &  23.66 $\pm$  0.17  &   30  &     2024 Mar 29 (F)\\
$J132514.41+034006.0$  &  25.92 $\pm$  0.22  &  23.90 $\pm$  0.07  &  23.80 $\pm$  0.10  &   30  &     2024 Mar 30 (F)\\
$J104800.36+011856.4$  &  25.56 $\pm$  0.14  &  23.62 $\pm$  0.03  &  23.57 $\pm$  0.10  &   50  &     2023 Jan 11 (F)\\
$J021103.93+015516.9$  &  25.33 $\pm$  0.09  &  23.28 $\pm$  0.03  &  23.20 $\pm$  0.06  &   30  &     2024 Nov 08 (F)\\
$J135059.71+023717.3$  &  25.99 $\pm$  0.18  &  23.88 $\pm$  0.05  &  23.87 $\pm$  0.12  &   30  &     2024 Mar 30 (F)\\
$J014056.95-005048.7$  &  25.61 $\pm$  0.25  &  23.54 $\pm$  0.08  &  23.98 $\pm$  0.15  &   50  &     2021 Dec 23 (F)\\
$J131024.32+005002.0$  &  27.43 $\pm$  0.69  &  23.76 $\pm$  0.05  &  23.65 $\pm$  0.10  &   50  &     2024 Mar 28 (F)\\
$J003554.15+030807.3$  &  24.11 $\pm$  0.04  &  22.65 $\pm$  0.02  &  22.44 $\pm$  0.03  &   15  &     2021 Nov 27 (F)\\
$J120542.50+041430.1$  &  25.70 $\pm$  0.11  &  23.69 $\pm$  0.06  &  24.13 $\pm$  0.16  &   30  &     2022 Dec 13 (F)\\
$J103825.94-004418.9$  &  26.21 $\pm$  0.20  &  23.82 $\pm$  0.04  &  23.66 $\pm$  0.08  &   30  &     2024 Mar 29 (F)\\
$J124359.28+015305.5$  &  25.86 $\pm$  0.33  &  23.78 $\pm$  0.05  &  23.80 $\pm$  0.10  &   30  &     2024 Mar 30 (F)\\
$J000928.93+001644.7$  &  25.32 $\pm$  0.09  &  23.75 $\pm$  0.06  &  23.93 $\pm$  0.12  &   30  &     2021 Dec 25 (F)\\
$J234434.18+022645.7$  &  24.97 $\pm$  0.08  &  23.36 $\pm$  0.04  &  23.53 $\pm$  0.10  &   30  &     2024 Nov 07 (F)\\
$J123934.04+025845.8$  &  25.99 $\pm$  0.23  &  23.88 $\pm$  0.05  &  23.85 $\pm$  0.08  &   30  &     2024 Mar 30 (F)\\
$J000440.93+010500.8$  &  25.35 $\pm$  0.09  &  23.73 $\pm$  0.05  &  23.70 $\pm$  0.09  &   30  &     2021 Nov 27 (F)\\
$J113411.21-001956.1$  &  25.33 $\pm$  0.09  &  23.63 $\pm$  0.05  &  24.24 $\pm$  0.18  &   60  &     2024 Apr 03 (O)\\
$J145945.34+415206.4$  &  23.85 $\pm$  0.08  &  22.34 $\pm$  0.04  &  22.56 $\pm$  0.07  &   15  &     2024 Apr 13 (O)\\
\hline\multicolumn{6}{c}{[\ion{O}{3}] Emitters}\\\hline
$J000526.30-001247.9$  &  25.42 $\pm$  0.10  &  26.36 $\pm$  0.54  &  23.61 $\pm$  0.09  &   20  &     2021 Dec 24 (F)\\
$J005743.52+044117.4$  &  25.20 $\pm$  0.14  &  $>$25.13   &  23.54 $\pm$  0.09  &   10  &     2021 Dec 24 (F)\\
$J090048.11-020909.6$  &  24.05 $\pm$  0.06  &  22.52 $\pm$  0.03  &  23.27 $\pm$  0.13  &   30  &     2024 Oct 22 (O)\\
$J023836.16-021027.8$  &  25.82 $\pm$  0.62  &  23.68 $\pm$  0.09  &  24.96 $\pm$  0.58  &   20  &     2021 Dec 26 (F)\\
$J020216.24+025309.4$  &  26.34 $\pm$  0.29  &  23.91 $\pm$  0.08  &  24.72 $\pm$  0.38  &   10  &     2021 Dec 24 (F)\\
\hline\multicolumn{6}{c}{Brown Dwarfs}\\\hline
$J002006.91+021707.3$  &  24.48 $\pm$  0.05  &  22.90 $\pm$  0.04  &  22.83 $\pm$  0.07  &   45  &     2024 Aug 10 (O)\\
$J004314.13+051600.4$  &  23.98 $\pm$  0.06  &  22.46 $\pm$  0.03  &  22.60 $\pm$  0.09  &   20  &     2024 Sep 06 (O)\\
$J004604.07+032818.6$  &  25.04 $\pm$  0.08  &  23.76 $\pm$  0.05  &  23.31 $\pm$  0.06  &   30  &     2021 Nov 25 (F)\\
$J005534.96+051051.0$  &  25.57 $\pm$  0.21  &  22.67 $\pm$  0.02  &  21.90 $\pm$  0.03  &   35  &     2024 Aug 12 (O)\\
$J011926.03+035801.2$  &  24.89 $\pm$  0.10  &  23.29 $\pm$  0.07  &  23.38 $\pm$  0.13  &  120  &     2024 Sep 06 (O)\\
$J013834.04+031801.3$  &  24.79 $\pm$  0.07  &  23.84 $\pm$  0.07  &  23.36 $\pm$  0.11  &   20  &     2021 Nov 26 (F)\\
$J014653.54+043154.6$  &  24.93 $\pm$  0.13  &  23.33 $\pm$  0.08  &  23.44 $\pm$  0.13  &  120  &     2024 Sep 07 (O)\\
$J015507.67+034557.2$  &  25.24 $\pm$  0.04  &  23.26 $\pm$  0.05  &  23.09 $\pm$  0.11  &   60  &     2024 Sep 11 (O)\\
$J021331.56-021215.3$  &  25.24 $\pm$  0.16  &  23.78 $\pm$  0.07  &  23.36 $\pm$  0.07  &   40  &  2021 Nov 26,27 (F)\\
$J022330.97+000216.7$  &  24.72 $\pm$  0.05  &  23.13 $\pm$  0.04  &  23.06 $\pm$  0.07  &   60  &     2024 Sep 15 (O)\\
$J023549.99-001719.5$  &  24.45 $\pm$  0.05  &  22.40 $\pm$  0.04  &  22.30 $\pm$  0.05  &   20  &     2024 Aug 12 (O)\\
$J083814.13+013741.8$  &  26.78 $\pm$  0.36  &  23.77 $\pm$  0.04  &  22.75 $\pm$  0.03  &   40  &     2021 Dec 24 (F)\\
$J084007.89+051351.7$  &  $>$25.04   &  22.74 $\pm$  0.07  &  21.92 $\pm$  0.06  &   20  &     2024 Nov 06 (F)\\
$J084427.33+022508.0$  &  26.00 $\pm$  0.22  &  23.21 $\pm$  0.04  &  22.20 $\pm$  0.02  &   70  &  2021 Dec 23,24 (F)\\
$J092701.24-012812.0$  &  27.09 $\pm$  0.63  &  23.65 $\pm$  0.10  &  22.97 $\pm$  0.07  &   40  &     2024 Nov 06 (F)\\
$J102059.04+030252.2$  &  25.41 $\pm$  0.12  &  23.20 $\pm$  0.04  &  22.65 $\pm$  0.05  &   40  &     2024 Nov 07 (F)\\
$J110813.13+012353.8$  &  24.89 $\pm$  0.06  &  23.18 $\pm$  0.04  &  22.93 $\pm$  0.06  &   40  &     2024 Nov 07 (F)\\
$J110931.21+004900.2$  &  25.25 $\pm$  0.09  &  23.58 $\pm$  0.03  &  23.29 $\pm$  0.08  &   60  &     2024 Mar 15 (O)\\
$J114838.82-021153.5$  &  24.61 $\pm$  0.08  &  22.16 $\pm$  0.02  &  21.21 $\pm$  0.02  &   30  &     2023 Mar 18 (O)\\
$J133642.85+435628.7$  &  26.79 $\pm$  0.38  &  25.79 $\pm$  0.42  &  21.76 $\pm$  0.02  &   30  &     2022 Apr 24 (F)\\
$J140416.94+045001.5$  &  25.53 $\pm$  0.14  &  22.85 $\pm$  0.05  &  21.79 $\pm$  0.03  &   60  &     2025 Mar 22 (O)\\
$J141220.77-012001.0$  &  27.43 $\pm$  0.53  &  26.05 $\pm$  0.37  &  23.85 $\pm$  0.09  &   50  &     2022 Dec 26 (F)\\
$J220451.87+063342.3$  &  $>$26.05   &  $>$24.55   &  22.13 $\pm$  0.05  &   60  &     2024 Sep 06 (O)\\
$J220627.21+064136.4$  &  23.76 $\pm$  0.07  &  22.26 $\pm$  0.06  &  22.39 $\pm$  0.14  &   16  &     2024 Aug 10 (O)\\
$J221336.02-000333.2$  &  24.54 $\pm$  0.05  &  21.69 $\pm$  0.01  &  20.66 $\pm$  0.01  &   30  &     2024 Aug 10 (O)\\
$J221753.95+002342.3$  &  26.72 $\pm$  0.29  &  25.12 $\pm$  0.22  &  23.25 $\pm$  0.05  &   50  &     2021 Nov 27 (F)\\
$J222736.05+042920.1$  &  $>$25.94   &  25.13 $\pm$  0.24  &  22.94 $\pm$  0.06  &   30  &     2021 Nov 26 (F)\\
$J224011.93-012705.4$  &  26.84 $\pm$  0.75  &  22.52 $\pm$  0.02  &  21.71 $\pm$  0.04  &   60  &     2024 Oct 31 (O)\\
$J224336.11+031740.5$  &  26.81 $\pm$  0.60  &  25.00 $\pm$  0.22  &  22.95 $\pm$  0.09  &   40  &     2021 Nov 26 (F)\\
$J234812.70+053951.8$  &  24.42 $\pm$  0.13  &  22.83 $\pm$  0.06  &  23.21 $\pm$  0.15  &   45  &     2024 Sep 13 (O)\\
\hline\multicolumn{6}{c}{Passive galaxies}\\\hline
$J004219.46+045530.1$  &  24.54 $\pm$  0.07  &  22.84 $\pm$  0.07  &  23.66 $\pm$  0.13  &   50  &     2024 Nov 08 (F)\\
$J004314.81+051632.5$  &  25.10 $\pm$  0.16  &  23.12 $\pm$  0.07  &  23.01 $\pm$  0.09  &   30  &     2024 Nov 08 (F)\\
$J011354.43+060446.8$  &  24.50 $\pm$  0.13  &  22.81 $\pm$  0.06  &  22.64 $\pm$  0.08  &   50  &     2024 Nov 09 (F)\\
$J011949.18+055925.0$  &  24.34 $\pm$  0.07  &  22.75 $\pm$  0.05  &  22.67 $\pm$  0.07  &   30  &     2024 Nov 06 (F)\\
$J012148.05+053026.6$  &  24.72 $\pm$  0.08  &  23.21 $\pm$  0.06  &  23.19 $\pm$  0.09  &   40  &     2024 Nov 09 (F)\\
$J012251.39+045000.0$  &  24.83 $\pm$  0.07  &  23.33 $\pm$  0.06  &  23.36 $\pm$  0.08  &   40  &     2024 Nov 09 (F)\\
$J012703.23+055558.9$  &  24.63 $\pm$  0.10  &  23.07 $\pm$  0.06  &  23.04 $\pm$  0.09  &   30  &     2024 Nov 08 (F)\\
$J012300.92+024904.0$  &  24.51 $\pm$  0.06  &  23.19 $\pm$  0.04  &  22.86 $\pm$  0.06  &   40  &     2021 Dec 23 (F)\\
$J012821.85+040331.3$  &  24.60 $\pm$  0.06  &  23.50 $\pm$  0.05  &  23.18 $\pm$  0.08  &   60  &     2024 Aug 13 (O)\\
$J104110.52+033845.3$  &  $>$26.11   &  25.17 $\pm$  0.19  &  22.79 $\pm$  0.08  &   30  &     2022 Dec 13 (F)\\
$J121643.52+043806.1$  &  26.68 $\pm$  0.32  &  23.66 $\pm$  0.08  &  23.16 $\pm$  0.08  &   50  &     2024 Mar 28 (F)\\
$J144147.28+042329.5$  &  23.58 $\pm$  0.02  &  21.97 $\pm$  0.02  &  22.37 $\pm$  0.04  &   15  &     2024 Jun 30 (O)\\
$J162307.07+421418.4$  &  24.35 $\pm$  0.04  &  22.85 $\pm$  0.04  &  22.68 $\pm$  0.05  &   30  &     2024 Apr 13 (O)\\
$J220252.90+065600.4$  &  24.54 $\pm$  0.07  &  22.99 $\pm$  0.06  &  23.05 $\pm$  0.13  &   50  &     2024 Nov 08 (F)\\
\enddata
\tablecomments{
Coordinates are at J2000.0, and magnitude lower limits are placed at $5\sigma$ significance. 
All magnitudes were taken from the HSC-SSP S23B DR, and were corrected for Galactic extinction.
The instrument (Inst) ``F" and ``O" denote Subaru/FOCAS and GTC/OSIRIS, respectively.}
\end{deluxetable*}

\end{document}